%% file: main.tex
\documentclass[screen, acmsmall]{acmart}
\AtBeginDocument{%
  }

\setcopyright{acmlicensed}
\acmDOI{XXXXXXX.XXXXXXX}
\acmConference[Conference acronym 'XX]{Make sure to enter the correct
  conference title from your rights confirmation emai}{June 03--05,
  2018}{Woodstock, NY}
\acmISBN{978-1-4503-XXXX-X/18/06}

\usepackage{graphicx}
\usepackage{listings}
\usepackage{caption}
\usepackage{subcaption}
\usepackage{tabularx}
\usepackage{comment}
\usepackage{balance}
\usepackage{colortbl}
\usepackage{rotating}
\usepackage{mdframed}
\usepackage{xspace}
\usepackage[linesnumbered,ruled,vlined]{algorithm2e}
\usepackage[most]{tcolorbox}
\usepackage{multirow}
\usepackage{amsmath}
\usepackage{graphicx}
\usepackage{subcaption}

\newboolean{showcomments}
\setboolean{showcomments}{true}         
\ifthenelse{\boolean{showcomments}}
  {\newcommand{\nb}[2]{
  \fbox{\bfseries\sffamily\scriptsize#1}
     {\sf\small$\blacktriangleright$\textit{\textcolor{red}{#2}}$\blacktriangleleft$}
   }
  }
  {\newcommand{\nb}[2]{}
   
  }

\newcommand{\COMMENT}[1]{}
\newcommand\name{MIST\xspace}

\begin{document}
\nocite{*}

\title{Detecting Trojaned DNNs via Spectral Regression Analysis}

\author{Samuele Pasini}
\email{samuele.pasini@usi.ch}
\orcid{0000-0002-7900-3727}
\affiliation{%
  \institution{Universit\`a della Svizzera Italiana}
  \city{Lugano}
  \country{Switzerland}
}

\author{Jinhan Kim}
\email{jinhan.kim@usi.ch}
\orcid{0000-0002-0140-7908}
\affiliation{%
  \institution{Universit\`a della Svizzera italiana}
  \city{Lugano}
  \country{Switzerland}
}

\author{Paolo Tonella}
\email{paolo.tonella@usi.ch}
\orcid{0000-0003-3088-0339}
\affiliation{%
  \institution{Universit\`a della Svizzera italiana}
  \city{Lugano}
  \country{Switzerland}
}

\renewcommand{\shortauthors}{Pasini et al.}

\begin{abstract}
Modern DNNs are repeatedly fine-tuned to incorporate new data and functionality. This evolutionary workflow introduces a security risk when updated data cannot be fully trusted, as adversaries may implant Trojans during fine-tuning.
We present \name, a Trojan detection approach that analyzes how a model’s internal representations change during fine-tuning. Rather than attempting to reconstruct trigger conditions, \name characterizes benign model evolution using pre-activation spectra and flags updates whose spectral deviations are inconsistent with this reference. This framing treats Trojan detection as a regression problem over model updates. An empirical evaluation across four datasets and eight Trojan attacks shows that spectral distances reliably distinguish Trojaned updates from clean fine-tuning. \name outperforms state-of-the-art detection accuracy after a single update, without requiring any knowledge about the poisoned data or the trigger, and remains effective under multi-step benign evolution, with graceful and bounded degradation. These results indicate that spectral evolution provides a stable and assumption-light signal for detecting malicious model updates.
\end{abstract}


\maketitle

\input{introduction}
\input{background}
\input{related}
\input{approach}

\input{empirical}
\input{results}
\input{threats}
\input{conclusion}
\section*{Conflict of Interest and Data Availability}

The implementations and the source code are publicly available in a GitHub repository\footnote{\url{https://github.com/PasiniSamuele/MIST}}.
For storage limit reasons, the checkpoints are not stored in any public repository, and they are available upon request

\begin{acks}
This work is funded by the European Union's Horizon Europe research and innovation programme under the project Sec4AI4Sec, grant agreement No 101120393.
\end{acks}

\bibliographystyle{IEEEtran}
\balance
\bibliography{onlytaxbib, bib}

\newpage

\end{document}

%% file: introduction.tex
\section{Introduction}
\label{sec:intro}

Deep Neural Networks (DNNs) are increasingly deployed as long-lived software artifacts in safety- and security-critical systems, ranging from autonomous driving to biometric authentication~\cite{he2016deep, liu2015faceattributes}. In such settings, models are rarely trained once and frozen. Instead, they are continuously updated through fine-tuning to incorporate new data, adapt to evolving environments, or support downstream tasks~\cite{de2021continual}. From a software engineering perspective, this practice turns DNNs into evolving programs whose correctness and trustworthiness must be reassessed after each update~\cite{zhang2020machine}.

This evolutionary workflow introduces a security risk when the data used for fine-tuning cannot be fully trusted~\cite{gu2017badnets, chen2017targeted}. Update data may originate from user feedback, automated collection pipelines, or third-party sources, and is often only partially inspected. An adversary can exploit this situation by injecting poisoned samples into the update data, thereby implanting a Trojan (or backdoor) into the fine-tuned model. A Trojaned model preserves its accuracy on clean inputs but exhibits attacker-chosen behavior when a specific trigger condition is met. Since this behavior is encoded implicitly in the model parameters rather than in explicit program logic, it is difficult to expose through manual inspection or standard testing.

Detecting such Trojans is challenging because the malicious behavior is activated by conditions that are unknown to the defender and may not manifest as identifiable input patterns. Triggers can be imperceptible, input-dependent, or embedded in the feature space, making them difficult to expose through direct input analysis. Consequently, many existing defenses analyze a trained model in isolation and attempt to recover or approximate potential trigger conditions~\cite{wang2019neural, liu2019abs, shen2021backdoor}. The effectiveness of these approaches therefore depends on assumptions about how the Trojan manifests at the input level, which may not hold uniformly across different attack strategies~\cite{wang2022rethinking}.

In this work, we take a different perspective. Rather than reasoning about inputs or trigger patterns, we analyze the internal state of the model. Our detection is based on how a model’s internal representations change during fine-tuning. Specifically, we focus on \textit{pre-activation spectra}, which summarize layer-wise pre-activations as probability distributions and provide a stable characterization of learned representations. This perspective naturally aligns with the assumption that models evolve over time. When a new model version is produced via fine-tuning, a trusted predecessor remains available for comparison. Under this assumption, Trojan detection reduces to assessing whether the internal changes introduced by the update are consistent with benign adaptation, or whether they indicate a malicious modification. No assumptions are made about the trigger’s form, location, or visibility in the input space.

\begin{figure*}[t]
  \centering
  \begin{subfigure}{0.44\linewidth}
    \centering
    \includegraphics[width=\linewidth]{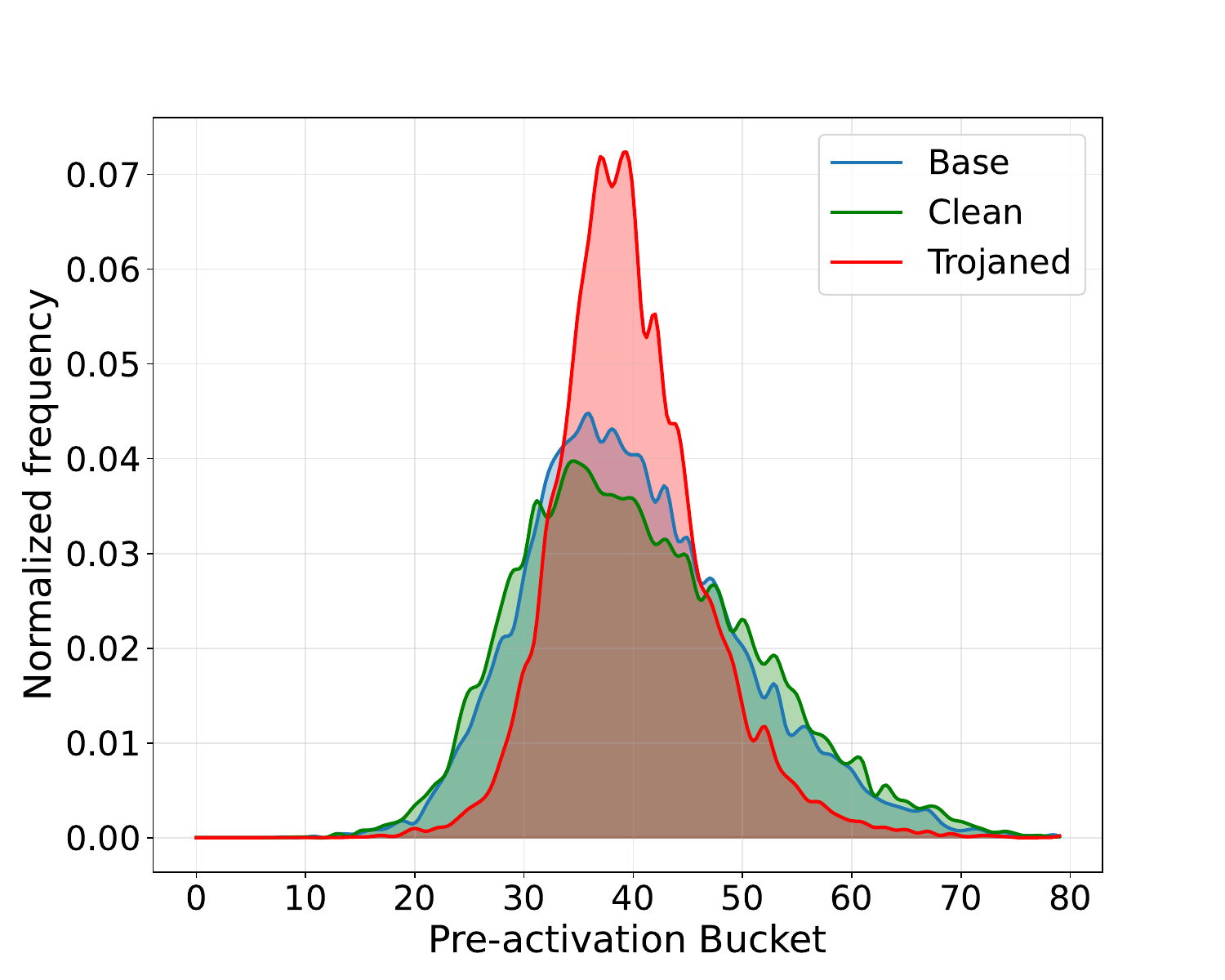}
    \caption{CIFAR-10}
  \end{subfigure}
  \hspace{0.01\linewidth}
  \begin{subfigure}{0.44\linewidth}
    \centering
    \includegraphics[width=\linewidth]{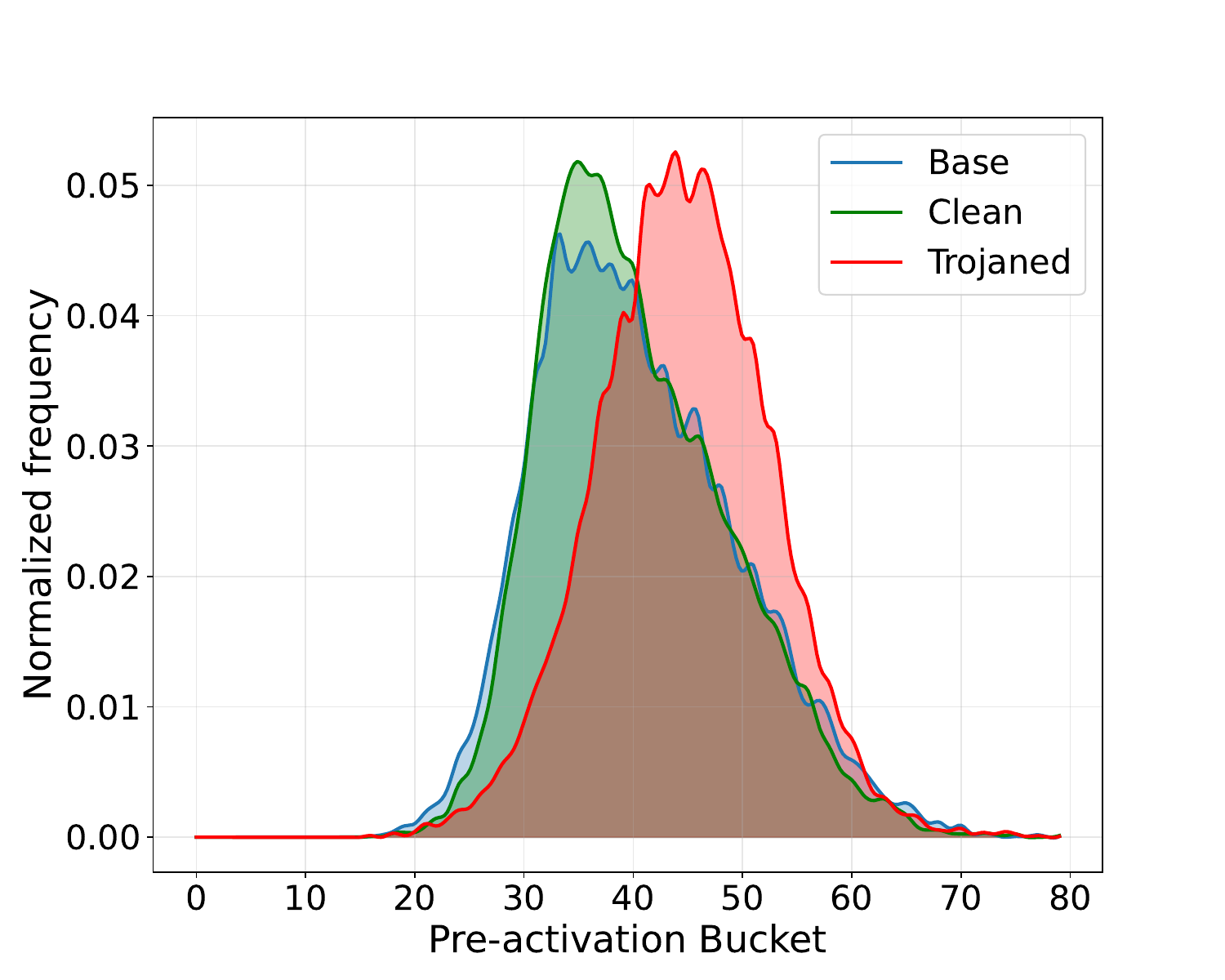}
    \caption{SVHN}
  \end{subfigure}

  \caption{Spectral differences induced by benign (Clean) and malicious (Trojaned) fine-tuning. The figure shows pre-activation spectra extracted from base models, cleanly fine-tuned models, and Trojaned models produced by a BadNet attack~\cite{gu2017badnets} on CIFAR-10 (on the left) and on SVHN (on the right). While benign fine-tuning induces limited spectral deviation from the base model, Trojan insertion results in substantially larger spectral changes. \name exploits this discrepancy to distinguish Trojaned updates from benign evolution.} \label{fig:spectra_distribution}
\end{figure*}

We instantiate this idea in \name (\underline{M}odel \underline{I}ncremental \underline{S}pectra \underline{T}racking), a Trojan detection technique that treats model updates as regression events and characterizes fine-tuning through spectral changes of internal representations. The underlying intuition is illustrated in Figure~\ref{fig:spectra_distribution}. Benign fine-tuning induces limited and structured spectral changes relative to a clean base model, whereas Trojan insertion introduces deviations that are markedly larger and qualitatively different. These anomalous spectral shifts provide a discriminative signal that persists even when the model’s external behaviour on clean inputs remains unchanged. Given a clean reference checkpoint, \name first learns a reference distribution of spectral differences induced by benign fine-tuning. It then flags an updated model as Trojaned if its spectral deviation from the reference exceeds what can be explained by benign evolution. This design makes \name naturally suited to deployment settings in which models are continuously updated and validated before redeployment.

We evaluate \name through three empirical studies. First, we validate the underlying assumption that spectral changes induced by Trojan insertion are separable from those caused by benign fine-tuning. Second, we assess the effectiveness of \name as a Trojan detector after a single fine-tuning step, comparing it against three state-of-the-art detection techniques across four datasets and eight Trojan attacks. Third, we examine robustness under multi-step benign evolution, analyzing how detection performance degrades as models drift further from the original reference checkpoint.

Our results show that spectral distances reliably distinguish Trojaned updates from clean ones. Across four datasets, \name achieves an average detection accuracy of 0.95 after a single fine-tuning step, outperforming three state-of-the-art Trojan detectors significantly. Under a multi-step evolution scenario, detection remains effective: although repeated fine-tuning introduces benign drift, accuracy degrades gracefully to 0.89 on average, with errors remaining bounded and predictable. These findings indicate that spectral evolution captures a stable signal of malicious modification that persists even as models evolve, without requiring assumptions about trigger structure or visibility.

In summary, this paper makes the following technical contributions:
\begin{itemize}
    \item We introduce a regression-based view of Trojan detection that frames malicious updates as anomalous deviations in spectral model evolution, rather than as trigger reconstruction problems.
    \item We propose \name, a practical Trojan detection technique based on pre-activation spectra that validates fine-tuned model updates against a clean reference checkpoint.
    \item We present a comprehensive empirical evaluation across four datasets, eight Trojan attacks, and three state-of-the-art detectors, outperforming state-of-the-art accuracy after single-step updates and robust performance under multi-step benign evolution.
\end{itemize}

The remainder of this paper is structured as follows. Section~\ref{sec:background} introduces background on Trojan attacks and spectral analysis. Section~\ref{sec:related} reviews related work on Trojan detection and model validation. Section~\ref{sec:approach} presents the design of \name and formalizes spectral regression for detecting malicious model updates. Section~\ref{sec:empirical} describes the experimental setup, datasets, and baselines. Section~\ref{sec:results} reports and analyzes the empirical results for our three research questions. Section~\ref{sec:threats} discusses threats to validity, and Section~\ref{sec:conclusion} concludes the paper.

%% file: background.tex
\section{Background}
\label{sec:background}

In this section, we provide some background on  Trojan attacks in DNNs. Moreover, we introduce the fundamental concept of  \textit{Spectral Analysis}, the core of the defense strategy of \name.

\subsection{Trojans Attacks}
A DNN classifier $\mathcal{M}$ models a function $\mathcal{M}: \mathcal{X} \mapsto \mathcal{Y}$, where the input space $\mathcal{X}\subseteq \mathbb{R}^{l}$ is mapped to a finite set of labels $\mathcal{Y}$ with cardinality $l$.
Given a dataset composed of input and label pairs $\mathcal{D} = \{(x_1, y_1), (x_2, y_2),...(x_n, y_n)\}$, $\mathcal{M}$ learns the mapping function by minimizing  $\frac{1}{n} \sum_{(x_i,y_i)\in \mathcal{D}} \mathcal{L} (\mathcal{M}(x_i),y_i)$, where $\mathcal{L}$ is the \textit{Loss Function}.

A Trojan (backdoor) attack embeds a hidden behaviour into a model, causing it to predict an attacker-chosen \emph{target} label for any input $x'$ containing a specific perturbation pattern, known as a \textit{trigger}, while preserving the model’s performance on clean, unmodified inputs.
Attacks of this nature may be introduced at the stage of training, often through the practice known as \textit{Data Poisoning}. More specifically, given the target label $\tau$ and the input space $\mathcal{X'}\subseteq X$ containing all the inputs with the trigger, an additional dataset $\mathcal{D}_p:\{(x_1', y_1'), (x_2', y_2'),...(x_n', y_n')\}$, where $\forall (x'_i, y'_i) \in \mathcal{D}_p$: $x'_i \in \mathcal{X'} \land y'_i = \tau$. Because of this addition, the minimized function is $\frac{1}{n} \sum_{(x_i,y_i)\in \mathcal{D}} \mathcal{L} (\mathcal{M}(x_i),y_i) + \frac{1}{m} \sum_{(x'_i,y'_i)\in \mathcal{D}_p} \mathcal{L} (\mathcal{M}(x'_i),y'_i)$. 
This objective accounts for both the malicious behaviour of the classifier happening when the input contains the trigger and the normal behaviour happening when the input does not contain the trigger.

This family of attacks is particularly dangerous because the victim has no access to $\mathcal{D}_p$ and is unaware of $\tau$; as a result, the malicious behaviour cannot be directly exposed, since the available test samples do not contain the trigger required to activate it.
Previous work has shown that Trojan attacks are possible for a wide variety of neural network architectures, including those for Computer Vision (CV)~\cite{gu2017badnets,barni2019new,chen2017targeted,li2021invisible,liu2018trojaning,liu2020reflection,nguyen2020input,nguyenwanet}, Natural Language Processing (NLP)~\cite{chen2021badnl,qi2021hidden}, Graph Neural Networks (GNNs)~\cite{xi2021graph,zhang2021backdoor} and Reinforcement Learning (RL)~\cite{kiourti2020trojdrl,wang2021backdoorl}. Triggers can be very simple,  for example, a small visual pattern such as a patch~\cite{gu2017badnets}. Still, they can also be much more sophisticated, entailing complex transformations, such as changes implemented through adversarial methods~\cite{li2021invisible}. When the trigger corresponds to a fixed perturbation applied directly in the input space, the attack is referred to as an \textit{input-space Trojan}. If the trigger modifies latent features of the input, it is considered a \textit{feature-space Trojan}. During the years, researchers focused on building more effective trojans that could  evade the proposed defensive methods~\cite{li2021invisible,liu2020reflection}.

\subsection{Spectral Analysis}

DNNs compute predictions through a sequence of layer-wise transformations that produce intermediate activation values. For a classifier $\mathcal{M} : \mathcal{X} \rightarrow \mathcal{Y}$ and an input $x \in \mathcal{X}$, each layer $\ell$ computes a vector of pre-activations $z^{(\ell)}(x)$, followed by a non-linear transformation that yields the corresponding activations. These activation values encode the internal representation learned by the network. Previous research has shown that individual neuron activations are highly unstable across retraining runs~\cite{humbatova2024spectral}. Even when models share the same architecture, training data, and comparable accuracy, the values of individual activation units may differ substantially. This instability limits the usefulness of neuron-level activations as a basis for characterizing model behavior or comparing different model instances.

To address this issue, prior work has proposed analyzing DNNs at the level of activation distributions rather than individual units. Humbatova et al.~\cite{humbatova2024spectral} introduced \textit{spectral analysis} as a way to characterize DNNs through the distribution of their activation values. The central idea is that, while individual activations vary, the aggregate distribution of activation values within a layer tends to be more stable and more representative of the features learned by the model.

In spectral analysis, activation values are summarized using a \textit{spectrum}, defined as a discretized probability distribution over activation magnitudes. Since activations are continuous, values are first normalized and then partitioned into a fixed number of bins. A histogram is constructed by counting how many activation values fall into each bin and is subsequently normalized to obtain a probability distribution. This representation abstracts away from individual neurons while preserving the overall activation pattern produced by the network. Although originally defined over post-activation values, the same construction can be applied to pre-activations.
Spectral representations have been shown to capture meaningful information about model behavior. Humbatova et al.~\cite{humbatova2024spectral} demonstrated that spectra can be used to cluster models affected by similar fault types and to predict fault categories with reasonable accuracy. These results suggest that spectral properties of activations provide a stable and informative summary of a model’s internal behavior, complementing traditional performance-based metrics.

In this paper, spectral analysis serves as a building block for comparing successive versions of a model as it evolves over time. By summarizing internal activations as distributions, spectra enable robust comparisons between model checkpoints that are resilient to retraining noise yet sensitive to systematic changes in learned representations. This makes spectral analysis particularly well-suited for assessing whether the internal changes induced by fine-tuning are consistent with benign model evolution or indicative of anomalous, potentially malicious behavior.

%% file: related.tex
\section{Related Work}
\label{sec:related}

Trojan attacks challenge the traditional notion of testing and validation for DNNs: the malicious behavior is latent, rarely exercised by standard test inputs, and often decoupled from observable failures on clean data. Existing defenses address this challenge under different assumptions about how models are produced, deployed, and updated. These assumptions determine when and how a defense can be applied.

A group of defenses assumes that the adversary can only poison the training data, but does not control the training process itself~\cite{tran2018spectral,tang2021demon,levine2020deep,zhang2022bagflip,pal2024backdoor}. In this setting, defenses either attempt to identify and remove poisoned samples prior to training or aim to learn classifiers that remain robust when trained on partially corrupted data. While effective in controlled settings, these techniques require access to training corpora and are difficult to apply when data is continuously collected or only partially trusted. More importantly, they offer limited protection in supply-chain scenarios, where the defender receives a fully trained model or an updated checkpoint whose training data is unavailable.

Several approaches therefore focus on detecting Trojan behavior at inference time. Methods such as STRIP~\cite{gao2019strip}, SentiNet~\cite{chou2020sentinet}, and Februus~\cite{doan2020februus} operate on already trained and potentially poisoned models, aiming to identify malicious inputs before they activate a backdoor. These techniques rely on observable properties of triggers, for example, by perturbing inputs or analyzing spatial saliency. As a result, they are vulnerable to attacks in which triggers are distributed, imperceptible, or embedded in the feature space. In addition, the repeated transformations and checks required at runtime introduce non-trivial overhead during deployment. Another line of work assumes full access to a poisoned model and attempts to remove the backdoor prior to deployment. Proposed techniques include fine-tuning, distillation, and neuron pruning~\cite{wu2021adversarial,li2021neural,min2023towards,xie2023badexpert}. Although such sanitization approaches can reduce attack success rates, they implicitly assume that a Trojan is already known to exist and may degrade the model’s performance on clean inputs. From a testing perspective, modifying the model would require a downstream regression testing phase of the sanitized version, to check if it still preserves the intended behavior.

Most closely related to this paper are approaches that inspect a trained model offline to determine whether it contains a Trojan before deployment, without modifying it. These methods assume white-box access to the model and a set of clean inputs and attempt to expose malicious behavior through systematic analysis. A dominant strategy in this family is trigger reverse engineering: if a backdoor exists, it should be possible to synthesize an input pattern that reliably activates it. Neural Cleanse (NC)~\cite{wang2019neural} formalizes this idea by searching for minimal input perturbations that cause the model to predict a target label. Several subsequent techniques~\cite{liu2019abs, guo2019tabor, chen2019deepinspect, shen2021backdoor, wang2022rethinking} build on this paradigm. ABS~\cite{liu2019abs} analyzes neuron-level sensitivity through artificial stimulation, while FeatuRE~\cite{wang2022rethinking} extends reverse engineering to the feature space by identifying decision boundaries associated with backdoor behavior. 
Despite their differences, these approaches share a key assumption: that Trojan behavior can be revealed by explicitly reconstructing representative triggers. As attacks increasingly rely on input-dependent or distributed mechanisms, this assumption becomes increasingly fragile.
Tran et al.~\cite{tran2018spectral} identify triggers in the inputs using neuron activation spectra. While we share with them the use of spectra, the goal of the two approaches is quite different: identification of backdoors in the input data at inference time~\cite{tran2018spectral} vs detection of poisoned models (our approach). We do not rely on the availability of any poisoned data, as we determine if a model is likely poisoned or not considering only clean inputs.

Our work departs from trigger reverse engineering entirely. Rather than attempting to synthesize triggers or elicit malicious predictions, it analyzes whether the internal changes introduced during model updates are consistent with benign evolution. This shift in perspective enables Trojan detection without assumptions about the trigger’s form, location, or visibility. Under this scenario, offline inspection techniques that assume full model access~\cite{liu2019abs, wang2019neural, wang2022rethinking} are the most relevant points of comparison, and we therefore use them as baselines in our evaluation.

%% file: approach.tex
\section{Approach}
\label{sec:approach}

\begin{figure*}[htb]
  \centering
  \includegraphics[width=1\linewidth ,origin=c]{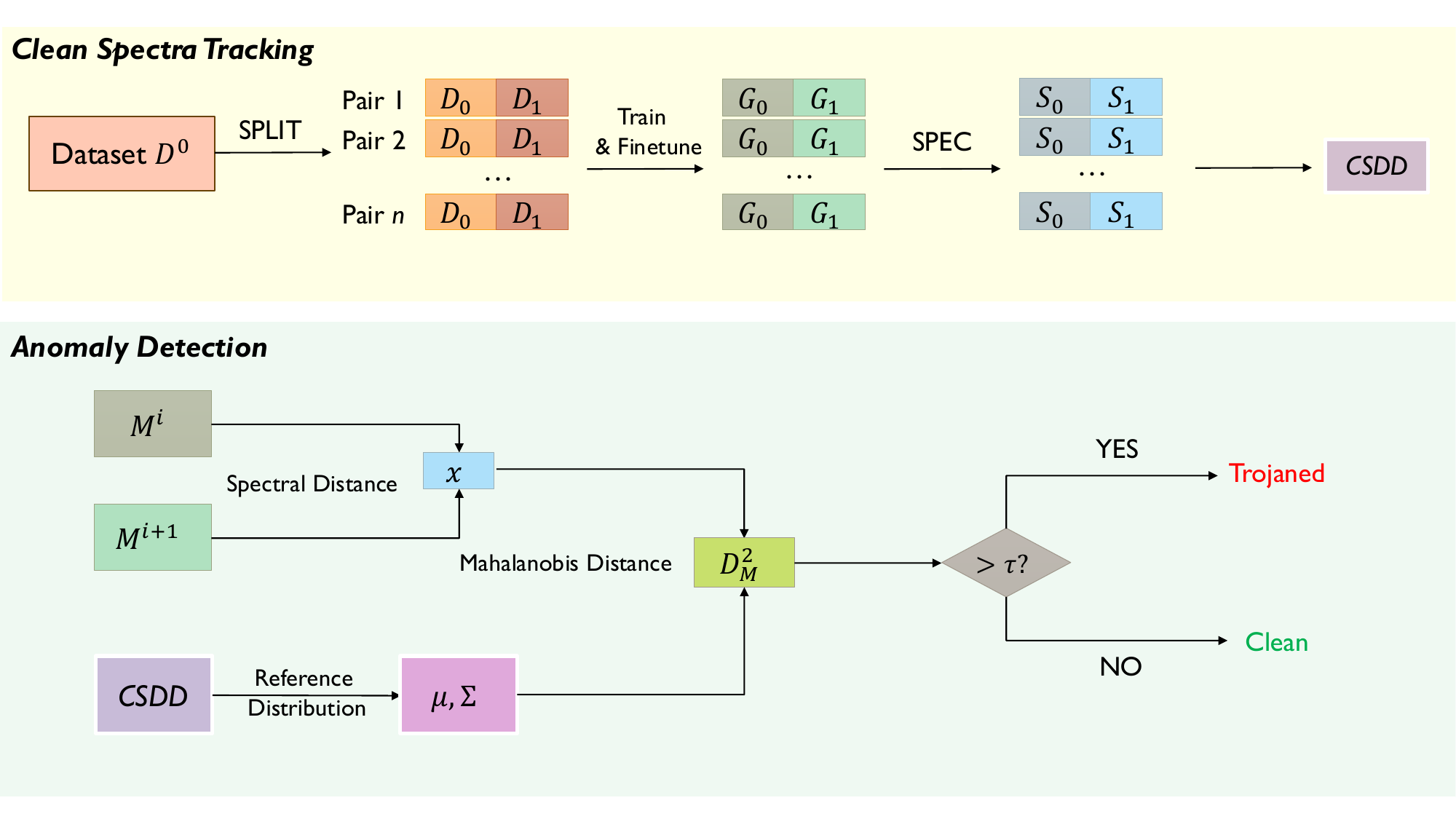}
  \caption{Overview of \name. The top part illustrates \textit{Clean Spectra Tracking}, which learns a reference distribution (CSDD) of spectral changes under benign model evolution. The bottom part shows the \textit{Anomaly Detection} phase, which compares a candidate model against this reference to determine whether the update is consistent with benign evolution or indicative of a Trojan.}\label{fig:approach}
\end{figure*}

This section presents \name, a method for detecting Trojaned neural networks by monitoring abnormal spectral changes induced during fine-tuning. The key insight behind \name is that, under benign conditions, fine-tuning a model on clean data induces regular and statistically stable changes in internal activation spectra. Trojan insertion, in contrast, perturbs these spectra in a systematic and detectable way. \name operationalizes this insight by (1) learning a reference distribution of clean spectral changes and (2) flagging deviations from this distribution as anomalies.

A summary of the approach is shown in Figure~\ref{fig:approach}. In the \textit{Clean Spectra Tracking} phase (Figure~\ref{fig:approach}, top), \name performs $n$ repeated clean training to fine tuning splits $D_0, D_1$, resulting in $n$ pairs. These pairs are used to train and then fine tune the model, resulting in $n$ trained model pairs $G_0, G_1$. Their associated spectra $S_0, S_1$ are then compared and the resulting distances define the Clean Spectra Distances Distribution (CSDD). In the \textit{Anomaly Detection} phase (Figure~\ref{fig:approach}, bottom), \name computes the spectral distance between two consecutive versions of the model $M^i, M^{i+1}$. 
It is important to notice that even when the evolved model $M^{i+1}$ is poisoned, we compute its spectrum on a clean test set, which means that \name makes no assumption on the availability of poisoned test samples or triggers. In fact, our hypothesis is that the spectral shift caused by the poison remains visible and measurable even when the poisoned model is exercised with clean test data only.
Based on mean $\mu$ and covariance $\Sigma$ of CSDD, \name computes the Mahalanobis distance $D^2_M$ of the observed spectral distance from the reference distribution CSDD. A threshold $\tau$ determines whether the model evolution is likely benign or potentially includes a Trojan.

Below, we first describe the threat model and application scenario addressed by \name, and then detail the two phases of the approach: \textit{Clean Spectra Tracking} and \textit{Anomaly Detection}.

\subsection{Threat Model and Application Scenario}
\label{sec:scenario}

\name targets what we call the \emph{Model Evolution} scenario, which arises when DNNs are treated as evolving software artifacts rather than static products. In practice, deployed models are repeatedly updated to incorporate new data, improve performance, or adapt to changing environments; from a testing perspective, each fine-tuning step produces a new program version whose correctness must be assessed before release. We assume that a defender initially deploys a model $\mathcal{M}^0$ trained on a fully trusted dataset $\mathcal{D}^0$. Over time, additional data $\mathcal{D}^1$ becomes available and is used to fine-tune the model, yielding a new checkpoint $\mathcal{M}^1$. While $\mathcal{D}^1$ is assumed to follow the same task distribution as $\mathcal{D}^0$, it is not fully trusted: the defender may lack the resources, authority, or time to exhaustively inspect all incoming data, and thus $\mathcal{D}^1$ may contain poisoned samples intended to implant a Trojan during fine-tuning. The defender lacks a direct oracle to determine whether $\mathcal{M}^1$ has been maliciously altered, since the attack may only manifest under rare or unknown trigger conditions.

To compensate for the absence of an explicit oracle, \name adopts a regression-testing perspective and evaluates whether the transition from $\mathcal{M}^0$ to $\mathcal{M}^1$ is consistent with benign model evolution. This comparison requires a controlled observation basis; for this reason, we assume that the defender can extract a small clean subset $\mathcal{D}_c^1 \subset \mathcal{D}^1$. Importantly, $\mathcal{D}_c^1$ is not used to retrain or repair the model, but solely as a testing instrument that enables trusted observations of internal model behavior. Such a subset is realistic in practice, as defenders often retain limited trusted data through manual vetting, provenance guarantees, or curated data sources even when the full dataset cannot be verified. Given a newly fine-tuned model $\mathcal{M}^{i+1}$ obtained from its predecessor $\mathcal{M}^i$ using an unknown dataset $\mathcal{D}^{i+1}$, the defender retains access to:
\begin{enumerate}
    \item the original clean training set $\mathcal{D}^0$, which establishes the baseline behavior of the model,
    \item a clean test set $\mathcal{D}_{test}^0$, which supports consistent cross-version comparisons, and
    \item a small clean subset $\mathcal{D}_c^{i+1}$, which serves as a trusted probe for observing internal model changes.
\end{enumerate}
\noindent
Using these artifacts, \name decides \emph{post hoc} whether the observed internal changes in $\mathcal{M}^{i+1}$ reflect expected, benign fine-tuning behavior or indicate a poisoned update.

\subsection{Algorithm}

Consistent with the model evolution scenario described in Section~\ref{sec:scenario}, \name treats each fine-tuned model as a new program version and evaluates whether the change from a previous checkpoint is consistent with benign evolution. To this end, the algorithm is divided into two phases: \textit{Clean Spectra Tracking}, which establishes a reference model of expected internal changes under clean updates (Figure~\ref{fig:approach}, top), and \textit{Anomaly Detection}, which checks whether a newly produced model deviates from this reference (Figure~\ref{fig:approach}, bottom).

\subsubsection{Clean Spectra Tracking}

\begin{algorithm}[htb]
\caption{Clean Spectra Tracking} \label{alg:cst}
\SetCommentSty{mycommfont}
\KwIn{
Clean training set $\mathcal{D}^0$, clean test set $\mathcal{D}_{test}^0$,
number of splits $n$, layer $\ell$, number of classes $C$, number of bins $B$
}
\KwOut{Clean Spectra Distances Distribution $\mathrm{CSDD} \in \mathbb{R}^{n \times C}$}
\BlankLine

\tcp{Initialize regression baseline for benign model evolution}
$\mathrm{CSDD} \gets$ empty $n \times C$ matrix\; \label{line:cst:init}

\For{$i \gets 1$ \KwTo $n$}{ \label{line:cst:outer}
    \tcp{Simulate a clean training-to-fine-tuning update}
    $(D_0, D_1) \gets \mathrm{SPLIT}(i, \mathcal{D}^0)$\; \label{line:cst:split}
    $G_0 \gets \mathrm{TRAIN}(\mathrm{INIT}(), D_0)$\; \label{line:cst:train}
    $G_1 \gets \mathrm{FINETUNE}(G_0, D_1)$\; \label{line:cst:finetune}
    
    \tcp{Measure internal change induced by the update}
    \For{$c \gets 0$ \KwTo $C-1$}{ \label{line:cst:inner}
        $S_0 \gets \mathrm{SPEC}(G_0, \mathcal{D}_{test}^0, \ell, c, B)$\; \label{line:cst:spec0}
        $S_1 \gets \mathrm{SPEC}(G_1, \mathcal{D}_{test}^0, \ell, c, B)$\; \label{line:cst:spec1}
        $\mathrm{CSDD}[i,c] \gets \| S_0 - S_1 \|_2$\; \label{line:cst:dist}
    }
}
\Return{$\mathrm{CSDD}$}\; \label{line:cst:return}
\end{algorithm}

Clean spectra tracking establishes a statistical baseline for how a model’s internal representations change under benign evolution. Algorithm~\ref{alg:cst} summarizes this process. The algorithm begins by initializing the CSDD, which stores the magnitude of internal changes observed across simulated clean updates (Line~\ref{line:cst:init}). To populate this baseline, \name repeatedly simulates \textit{benign training-to-fine-tuning} transitions using only trusted data (Line~\ref{line:cst:outer}). In each iteration, the clean training set $\mathcal{D}^0$ is partitioned into two equally sized subsets (Line~\ref{line:cst:split}); a model $G_0$ is trained from scratch on the first subset (Line~\ref{line:cst:train}) and then fine-tuned on the second subset to obtain $G_1$ (Line~\ref{line:cst:finetune}). The transition from $G_0$ to $G_1$ represents a controlled, fully benign update, analogous to updating a deployed model with newly acquired clean data.

To quantify how a model changes during such a clean update, \name compares the internal activation \textit{spectra} of the two consecutive checkpoints. Intuitively, a spectrum summarizes how strongly neurons in a given layer respond when the model predicts a specific class. 
In Algorithm~\ref{alg:cst}, this comparison is carried out after constructing each clean update pair (Lines~\ref{line:cst:train}--\ref{line:cst:finetune}).

Given a model $\mathcal{M}$, a dataset $\mathcal{D}$, a layer $\ell$, and a predicted class $c$, the operator $\mathrm{SPEC}(\mathcal{M}, \mathcal{D}, \ell, c, B)$ constructs the associated spectrum. When invoked at Lines~\ref{line:cst:spec0} and~\ref{line:cst:spec1}, this procedure first selects all inputs $x \in \mathcal{D}$ for which $\mathcal{M}$ predicts class $c$. For each selected input, it extracts the pre-activation vector $z^{(\ell)}(x)$ at layer $\ell$ and normalizes it by its absolute value to remove scale effects. The normalized values are then discretized into $B$ bins over the interval $[-1,1]$ and aggregated into a histogram. Finally, this histogram is normalized so that it forms a probability distribution, which we refer to as the activation spectrum of $\mathcal{M}$ at layer $\ell$ for class $c$.

Applying this procedure to the two checkpoints $G_0$ and $G_1$ yields a pair of per-class spectra, denoted $\tilde{S}_0^{(\ell,c),i}$ and $\tilde{S}_1^{(\ell,c),i}$. Algorithm~\ref{alg:cst} quantifies the internal change induced by the clean update by computing the $L_2$ distance between these distributions (Line~\ref{line:cst:dist}). Repeating this comparison for all classes (Line~\ref{line:cst:inner}) and across all simulated clean updates (Line~\ref{line:cst:outer}) produces the CSDD, an $n \times C$ matrix that captures the typical magnitude and variability of spectral changes under benign model evolution. This distribution serves as the regression baseline against which future model updates are evaluated.

\subsubsection{Anomaly Detection}

\begin{algorithm}[htb]
\caption{Anomaly Detection} \label{alg:anomaly}
\SetCommentSty{mycommfont}
\KwIn{
Model under test $\mathcal{M}^{i+1}$, Clean Model reference $\mathcal{M}^{i}$, 
Clean Spectra Distances Distribution $\mathrm{CSDD}$,
confidence level $\alpha$
}
\KwOut{Binary decision: \textsc{Clean} or \textsc{Poisoned}}
\BlankLine

\tcp{Fit reference distribution of benign spectral changes}
$\mu \gets$ mean of rows in $\mathrm{CSDD}$\; \label{line:ad:mean}
$\Sigma \gets$ covariance of rows in $\mathrm{CSDD}$\; \label{line:ad:cov}
$\widehat{\Sigma} \gets$ Ledoit--Wolf shrinkage of $\Sigma$\; \label{line:ad:lw}

\BlankLine
\tcp{Measure deviation of the new model update}
$x \gets \mathrm{SpectralDistance}(\mathcal{M}^{i+1}, \mathcal{M}^{i})$\; \label{line:ad:x}
$D_M^2 \gets (x - \mu)^{T} \widehat{\Sigma}^{-1} (x - \mu)$\; \label{line:ad:md}

\BlankLine
\tcp{Hypothesis test against benign evolution}
$\tau \gets \chi^2_{C}(\alpha)$\; \label{line:ad:thresh}
\eIf{$D_M^2 > \tau$}{ \label{line:ad:test}
    \Return \textsc{Poisoned}\;
}{
    \Return \textsc{Clean}\;
}
\end{algorithm}

Anomaly detection determines whether a newly fine-tuned model deviates from the expected pattern of benign evolution captured by the CSDD. Algorithm~\ref{alg:anomaly} summarizes this decision procedure, which is executed each time a new model version is produced and must be validated before deployment.

The algorithm begins by fitting a statistical reference model to the CSDD. Specifically, it computes the mean vector $\mu$ and covariance matrix $\Sigma$ over the clean spectral distance observations (Lines~\ref{line:ad:mean}--\ref{line:ad:cov}). Because the number of clean update samples is limited, the covariance matrix is regularized using the Ledoit--Wolf shrinkage estimator (Line~\ref{line:ad:lw}), which improves numerical stability and prevents ill-conditioned estimates.

Next, the algorithm measures how much the model under test $\mathcal{M}^{i+1}$ deviates from this clean reference $\mathcal{M}^{i}$ . Using the same procedure as in clean spectra tracking, it extracts a per-class spectral distance vector $x$ that summarizes the internal change induced by the latest update (Line~\ref{line:ad:x}). The deviation of this update is quantified by the squared Mahalanobis distance $D_M^2$ between $x$ and the clean reference distribution (Line~\ref{line:ad:md}), which accounts for correlations between class-wise spectral changes.
Under the assumption that clean spectral distances follow a multivariate Gaussian distribution, the squared Mahalanobis distance asymptotically follows a $\chi^2$ distribution with $C$ degrees of freedom. This enables anomaly detection to be framed as a statistical hypothesis test. Given a confidence level $\alpha$, the algorithm computes a detection threshold $\tau$ as the $\alpha$-quantile of the $\chi^2_C$ distribution (Line~\ref{line:ad:thresh}). If the observed deviation exceeds this threshold (Line~\ref{line:ad:test}), the update is flagged as anomalous and the model is classified as poisoned; otherwise, the update is deemed consistent with benign model evolution.

%% file: empirical.tex
\section{Empirical Study}
\label{sec:empirical}

This section presents the design of our empirical evaluation of \name as a Trojan detection technique under the model evolution scenario introduced in Section~\ref{sec:scenario}. The goal of our study is to assess whether the spectral changes exploited by \name provide a reliable and robust basis for distinguishing benign model updates from Trojaned ones, both in simple controlled settings and in realistic multi-step update scenarios.

\subsection{Research Questions}

Our evaluation is structured around three research questions, each targeting a distinct assumption or usage scenario of \name.
We begin with a preliminary experiment (RQ1) designed to isolate the core intuition behind \name:  Trojan insertion induces internal spectral changes that differ measurably from those caused by benign fine-tuning. To this end, we simplify the detection setting by directly comparing fine-tuned models against a clean reference checkpoint, without invoking the full anomaly detection pipeline. This experiment is not intended to demonstrate end-to-end detection performance, but rather to serve as a sanity check that validates the separability assumption on which \name is built.

\begin{itemize}
    \item \textbf{RQ1. Spectral Separability of Trojaned Models:}  
    \textit{Given a clean model checkpoint and a set of models fine-tuned from it using either clean or poisoned data, can spectral distances separate Trojaned models from clean ones?}
\end{itemize}

To answer this question, we compute the CSDD using clean training data, train a reference model $M^{0}$, and fine-tune it to obtain a clean updated model. Starting from the same checkpoint $M^{0}$, we then generate Trojaned models using the considered attacks. We compute spectral differences between all fine-tuned models and the reference model $M^{0}$, and use ROC--AUC to quantify how well spectral distances separate Trojaned models from clean ones.

Having established whether such separability exists, we then evaluate \name in its intended operational setting. The second research question assesses the effectiveness of \name as a Trojan detector after a single fine-tuning step, comparing its performance against state-of-the-art Trojan detection techniques. This setting reflects a common deployment scenario in which a model is updated once using newly acquired data and must be validated before redeployment.

\begin{itemize}
    \item \textbf{RQ2. Detection Effectiveness:}  
    \textit{How accurately does \name detect Trojaned models after a single fine-tuning step, compared to state-of-the-art Trojan detectors?}
\end{itemize}

To address this question, we execute the complete \name pipeline, including both Clean Spectra Tracking and the Anomaly Detection step, and evaluate its ability to correctly classify clean and Trojaned models under the previously specified assumptions. The same set of models is also analyzed using representative state-of-the-art Trojan detection techniques.

Finally, we examine the robustness of \name under repeated model evolution. In realistic deployments, models are often fine-tuned multiple times, and the effects of benign updates and malicious manipulations may accumulate across successive steps. The third research question investigates whether \name remains effective as the distance from the original clean checkpoint increases.

\begin{itemize}
    \item \textbf{RQ3. Robustness under Multi-step Evolution:}  
    \textit{How does the detection performance of \name evolve over multiple consecutive fine-tuning steps?}
\end{itemize}

To answer this question, we reuse the same CSDD computed for RQ2, but introduce an additional clean fine-tuning step before generating the models used for testing, thereby simulating multi-step model evolution. The detection procedure remains unchanged. Since this scenario is specific to the assumptions underlying \name, we do not compare against state-of-the-art detectors that are not designed for this setting; instead, we analyze performance relative to the single-step case studied in RQ2.

\subsection{Datasets and Models}

We evaluate \name on four widely used image classification datasets (see Table~\ref{tab:datasets}): CIFAR-10~\cite{krizhevsky2009learning}, SVHN~\cite{netzer2011reading}, GTSRB~\cite{stallkamp2011gtsrb}, and CelebA~\cite{liu2015faceattributes}. These datasets cover a range of domains, including natural objects, digit recognition, traffic signs, and human faces, allowing us to assess the generality of the proposed approach across diverse visual tasks. The CelebA dataset provides 40 binary facial attribute annotations rather than multi-class labels. Since this format is not directly suitable for multi-class classification, we follow the experimental configuration proposed by Salem et al.~\cite{salem2022dynamic}. Specifically, we select the three most balanced attributes (\textit{Heavy Makeup}, \textit{Mouth Slightly Open}, and \textit{Smiling}) and combine them to form an eight-class classification task.

\begin{table}[t]
\centering
\caption{Datasets used in our evaluation.}
\label{tab:datasets}
\begin{tabular}{llrrr}
\toprule
\textbf{Name} & \textbf{Domain} & \textbf{\# Classes} & \textbf{\# Training Samples} & \textbf{Input Size} \\
\midrule
CIFAR-10 & Natural objects & 10 & 50{,}000 & $3 \times 32 \times 32$ \\
SVHN & Street view house numbers & 10 & 73{,}257 & $3 \times 32 \times 32$ \\
GTSRB & Traffic signs & 43 & 39{,}209 & $3 \times 32 \times 32$ \\
CelebA & Human faces (attributes) & 8 & 202{,}599 & $3 \times 64 \times 64$ \\
\bottomrule
\end{tabular}
\end{table}

For all the datasets, we used as target DNN model ResNet18~\cite{he2016deep}. The CSDDs are computed using 15 training pairs. Following the recommendations by Humbatova et al.~\cite{humbatova2024spectral}, who extensively evaluated the properties of DNN spectra, we select a layer from the late ones for the computation of spectra. Specifically, in our work 
spectra are extracted from the last convolutional layer. In preliminary experiments, we also explored extracting spectra from additional layers; however, this consistently degraded detection performance, likely due to increased noise and reduced discriminative power.

\subsection{Trojan Attacks and Baselines Detectors}
\label{sec:attacks}

\begin{figure}[htbp]

    \centering

    \begin{minipage}{0.72\textwidth}
        \centering
        
        \begin{subfigure}{0.23\linewidth}
            \centering
            \includegraphics[width=\linewidth]{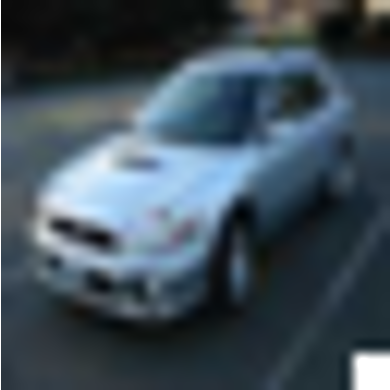}
            \caption{Badnet}
        \end{subfigure}\hfill
        \begin{subfigure}{0.23\linewidth}
            \centering
            \includegraphics[width=\linewidth]{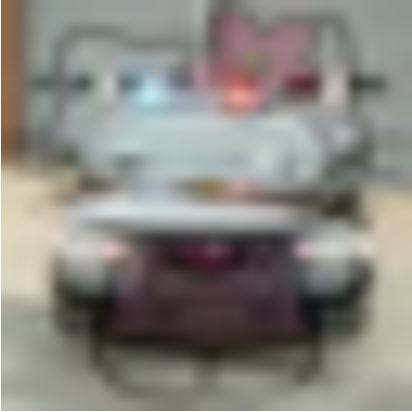}
            \caption{Blended}
        \end{subfigure}\hfill
        \begin{subfigure}{0.23\linewidth}
            \centering
            \includegraphics[width=\linewidth]{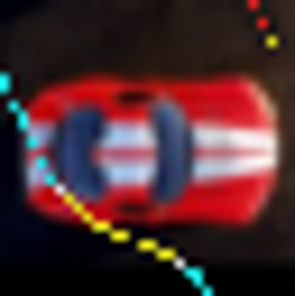}
            \caption{Input Aware}
        \end{subfigure}\hfill
        \begin{subfigure}{0.23\linewidth}
            \centering
            \includegraphics[width=\linewidth]{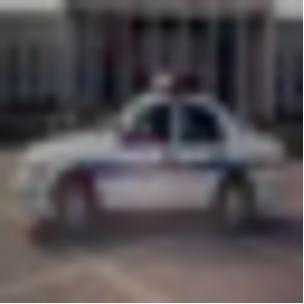}
            \caption{Refool}
        \end{subfigure}

        \medskip

        \begin{subfigure}{0.23\linewidth}
            \centering
            \includegraphics[width=\linewidth]{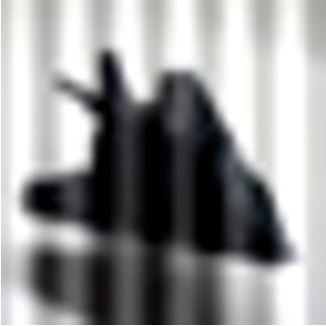}
            \caption{SIG}
        \end{subfigure}\hfill
        \begin{subfigure}{0.23\linewidth}
            \centering
            \includegraphics[width=\linewidth]{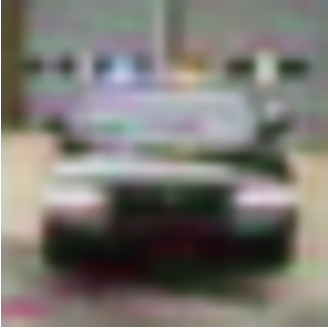}
            \caption{SSBA}
        \end{subfigure}\hfill
        \begin{subfigure}{0.23\linewidth}
            \centering
            \includegraphics[width=\linewidth]{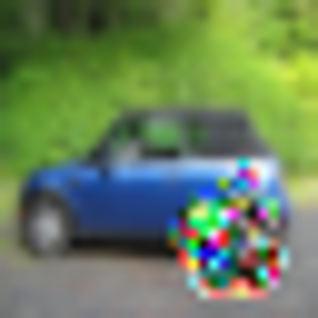}
            \caption{TrojanNN}
        \end{subfigure}\hfill
        \begin{subfigure}{0.23\linewidth}
            \centering
            \includegraphics[width=\linewidth]{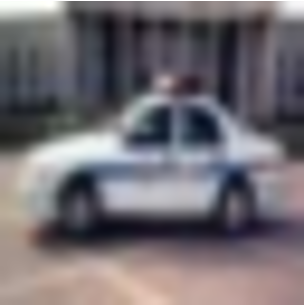}
            \caption{Wanet}
        \end{subfigure}
    \end{minipage}
    \caption{Example Trojaned samples of  CIFAR-10. For each attack, samples  with high trigger visibility have been selected; nevertheless, triggers produced by some attacks (e.g., Refool and Wanet) remain difficult to identify visually.}
\label{fig:attacks}

\end{figure}

We evaluated \name to detect 8 different types of established Trojan attacks. 
\textit{BadNet}~\cite{gu2017badnets} introduces a fixed, visible trigger stamped onto poisoned samples. \textit{Blended}~\cite{chen2017targeted} mixes a trigger pattern into images with controlled transparency, producing more subtle perturbations. \textit{Input-Aware}~\cite{nguyen2020input} generates input-dependent triggers that vary across samples, bypassing assumptions about static trigger patterns. \textit{Refool}~\cite{liu2020reflection} constructs natural-looking triggers by overlaying sample-specific reflection artifacts. \textit{SIG}~\cite{barni2019new} embeds sinusoidal perturbations that are visually inconspicuous yet easily learned. \textit{SSBA}~\cite{li2021invisible} similarly relies on imperceptible, sample-specific perturbations that avoid producing a consistent trigger. \textit{TrojanNN}~\cite{liu2018trojaning} implants a backdoor through model fine-tuning with synthetic triggers, enabling supply-chain attacks without access to clean training data. Finally, \textit{WaNet}~\cite{nguyenwanet} applies geometric warping transformations that remain imperceptible while activating the backdoor.
Examples of the trigger used to perform these Trojan attacks on the CIFAR-10 dataset are reported in Figure~\ref{fig:attacks}.

We compared \name with different SOTA DNN Trojan detectors described in Section~\ref{sec:related}: \textit{NC}~\cite{wang2019neural}, \textit{ABS}~\cite{liu2019abs}, \textit{FeatuRE}~\cite{wang2022rethinking}.
Although the techniques considered in this comparison work under a more relaxed set of assumptions (e.g., without requiring the availability of a clean starting point), \name has been proposed in a novel setup that was never considered before in the literature. Hence, we can only compare it with the set of Trojan detectors available from the literature.

\subsection{Evaluation Metrics}

We employ different evaluation metrics depending on the purpose of each experiment. For RQ1, we use the area under the ROC curve (ROC--AUC) to quantify how well spectral distances separate Trojaned models from clean ones, independent of any fixed detection threshold.

For RQ2 and RQ3, which evaluate \name as a binary Trojan detector, we report overall detection accuracy (Acc), defined as the proportion of correctly classified models, including both benign and Trojaned instances. 
We establish the detection threshold by selecting a confidence level $\alpha = 0.999$ and computing the threshold as $\tau = \chi^2_C(\alpha)$, which is the $\alpha$\textit{-quantile} of the chi-squared distribution of the Mahalanobis spectral distances.
To provide a more detailed view of detection behavior, we also report the full confusion matrix statistics: True Positives (TP), corresponding to Trojaned models correctly identified; False Positives (FP), benign models incorrectly flagged as Trojaned; False Negatives (FN), Trojaned models missed by the detector; and True Negatives (TN), benign models correctly recognized.

\subsection{Configurations}
All experiments are repeated ten times with different random dataset partitions to mitigate the effects of non-determinism due to data splitting and stochastic training. Reported results are averaged over these runs. Experiments are conducted on a machine equipped with an AMD EPYC 7742 64-Core CPU, an NVIDIA Tesla V100 GPU, and 512GB of RAM, running Ubuntu~20.04.6~LTS. Trojan attacks are implemented using the BackdoorBench framework\footnote{\url{https://github.com/SCLBD/BackdoorBench}} with default attack parameters.

%% file: results.tex
\section{Results}
\label{sec:results}

This section presents the results of our empirical evaluation with respect to the three research questions.

\begin{table}[t]
\centering
\caption{ROC--AUC of spectral distances for separating Trojaned and clean models across attacks and datasets.}
\label{tab:rq1}
\begin{tabular}{l rrrr}
\toprule
\textbf{Attack} & \textbf{CIFAR-10} & \textbf{SVHN} & \textbf{GTSRB} & \textbf{CelebA} \\
\midrule
BadNet      & 1.00 & 1.00 & 1.00 & 0.92 \\
Blended     & 1.00 & 0.97 & 1.00 & 0.99 \\
Input-Aware & 1.00 & 1.00 & 1.00 & 0.75 \\
Refool      & 1.00 & 0.93 & 1.00 & 0.74 \\
SIG         & 0.96 & 0.84 & 1.00 & 0.79 \\
SSBA        & 1.00 & 0.99 & 1.00 & 0.86 \\
TrojanNN    & 1.00 & 0.98 & 1.00 & 0.93 \\
WaNet       & 1.00 & 1.00 & 1.00 & 0.84 \\
\bottomrule
\end{tabular}
\end{table}

\subsection{Spectra Separability of Trojaned Models (RQ1)}

RQ1 examines the core assumption underlying \name: that Trojan insertion induces internal spectral changes that are systematically different from those caused by benign fine-tuning. To assess this assumption in isolation, we measure how well spectral distances separate models fine-tuned with poisoned data from those fine-tuned with clean data, using the clean reference checkpoint $M^{0}$ as baseline.

Table~\ref{tab:rq1} summarizes the resulting ROC--AUC values across datasets and attacks. Overall, spectral distances provide a strong separating signal: averaged over all attack--dataset combinations, the ROC--AUC reaches 0.95. For CIFAR-10, SVHN, and GTSRB, separability is near-perfect, with average AUCs of 0.99, 0.96, and 1.00, respectively. These results indicate that, for these datasets, Trojan-induced spectral changes are consistently larger than those introduced by benign fine-tuning. Separability is lower on CelebA, where the average AUC drops to 0.85. This degradation is expected, because CelebA is the hardest dataset to deal with, with a lot of images that differ between each other. Such variability makes the training and the poisoning process harder, thus separability is more difficult too. Nevertheless, even in this setting, spectral distances retain substantial discriminative power.

\begin{figure}[htbp]
    \centering

    \begin{subfigure}{0.24\textwidth}
        \centering
        \includegraphics[width=\linewidth]{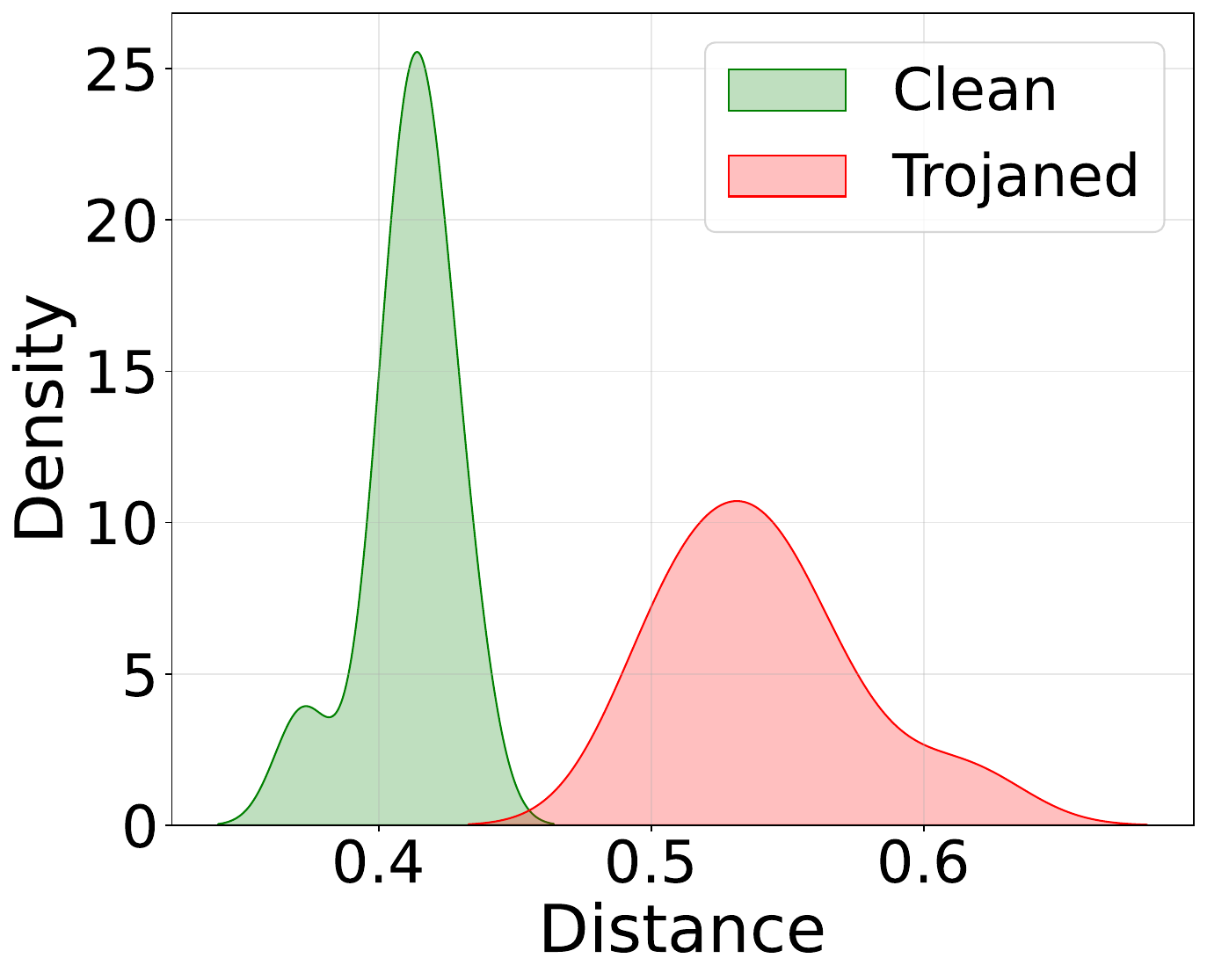}
        \caption{Badnet}
    \end{subfigure}\hfill
    \begin{subfigure}{0.24\textwidth}
        \centering
        \includegraphics[width=\linewidth]{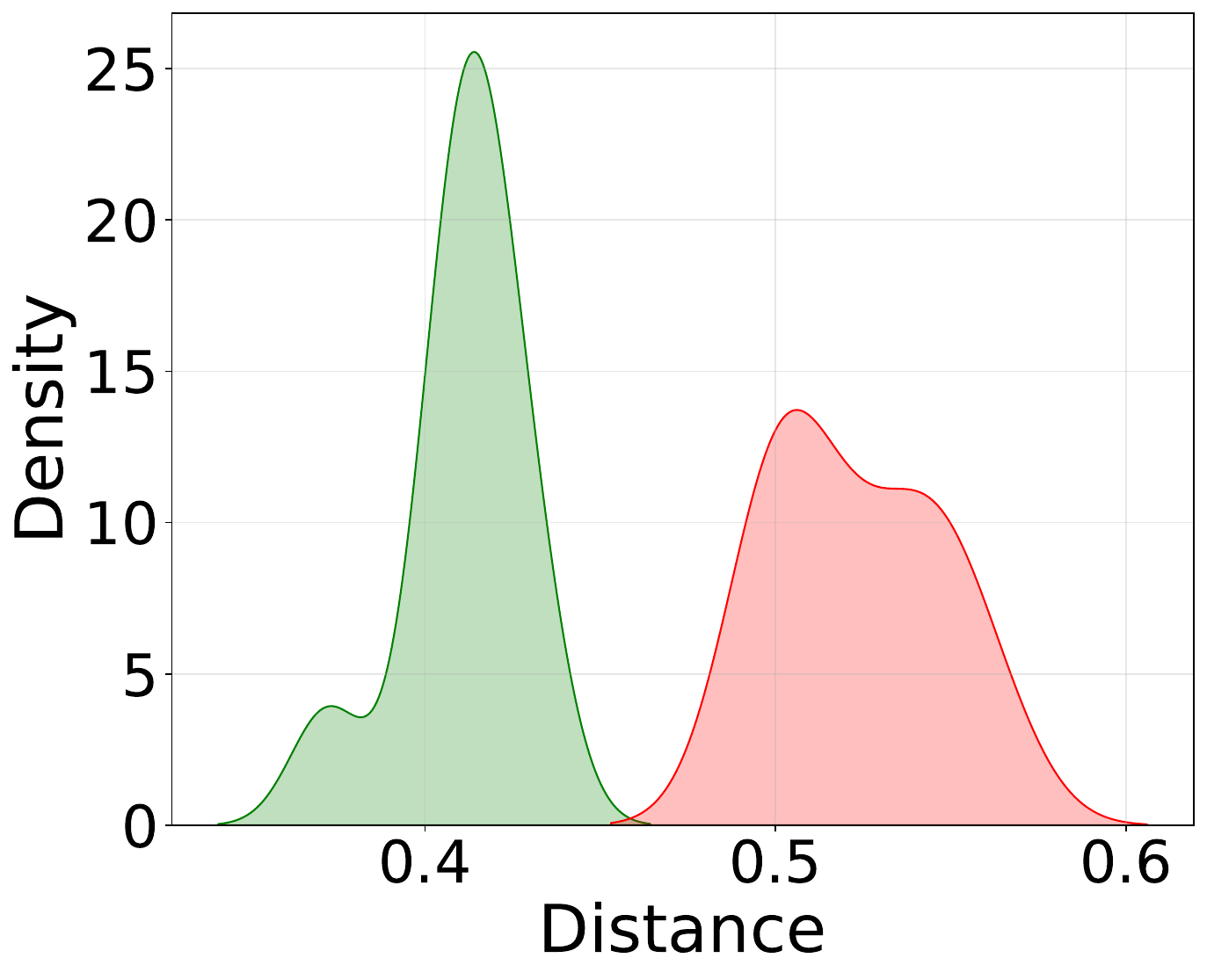}
        \caption{Blended}
    \end{subfigure}\hfill
    \begin{subfigure}{0.24\textwidth}
        \centering
        \includegraphics[width=\linewidth]{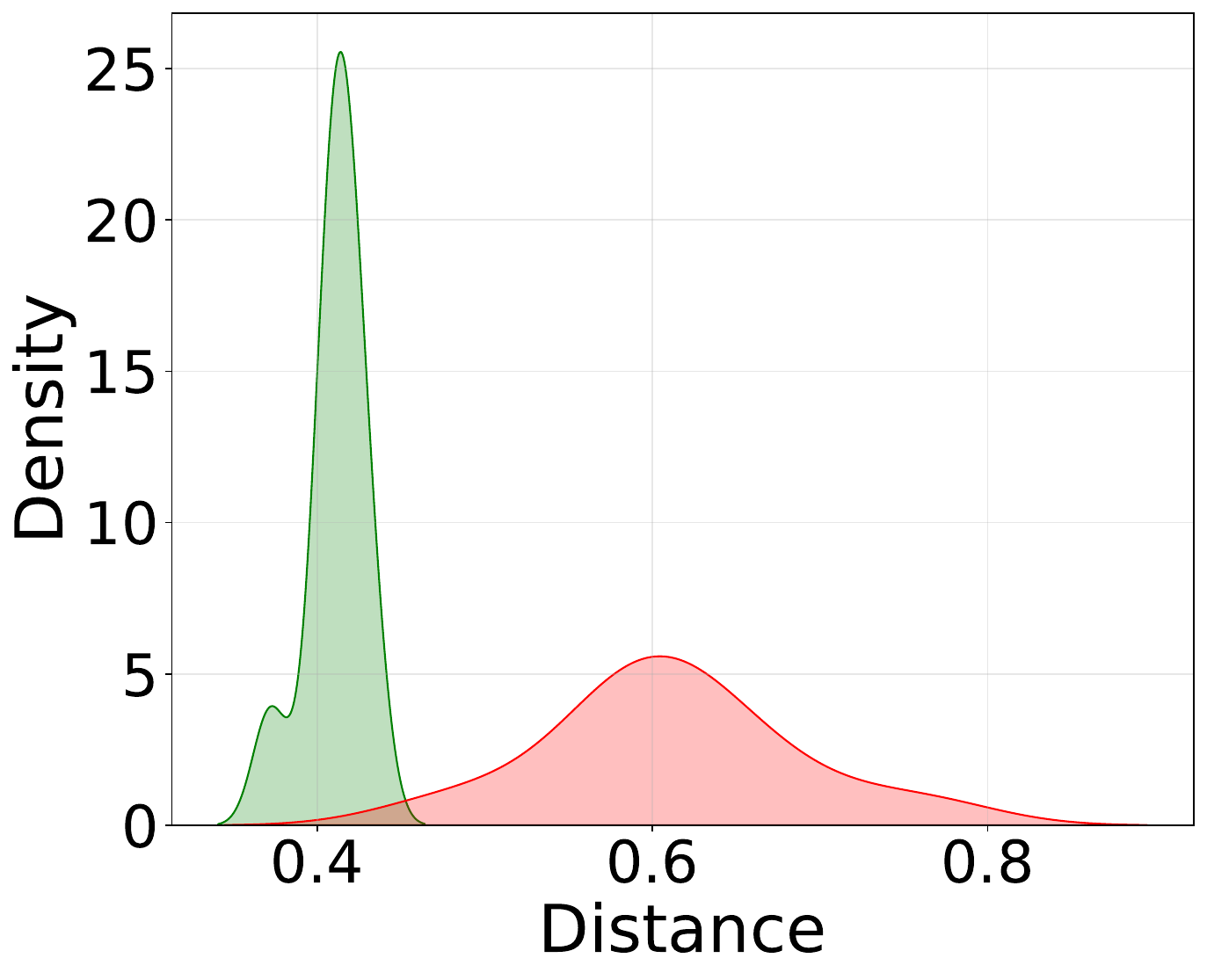}
        \caption{Input Aware}
    \end{subfigure}\hfill
    \begin{subfigure}{0.24\textwidth}
        \centering
        \includegraphics[width=\linewidth]{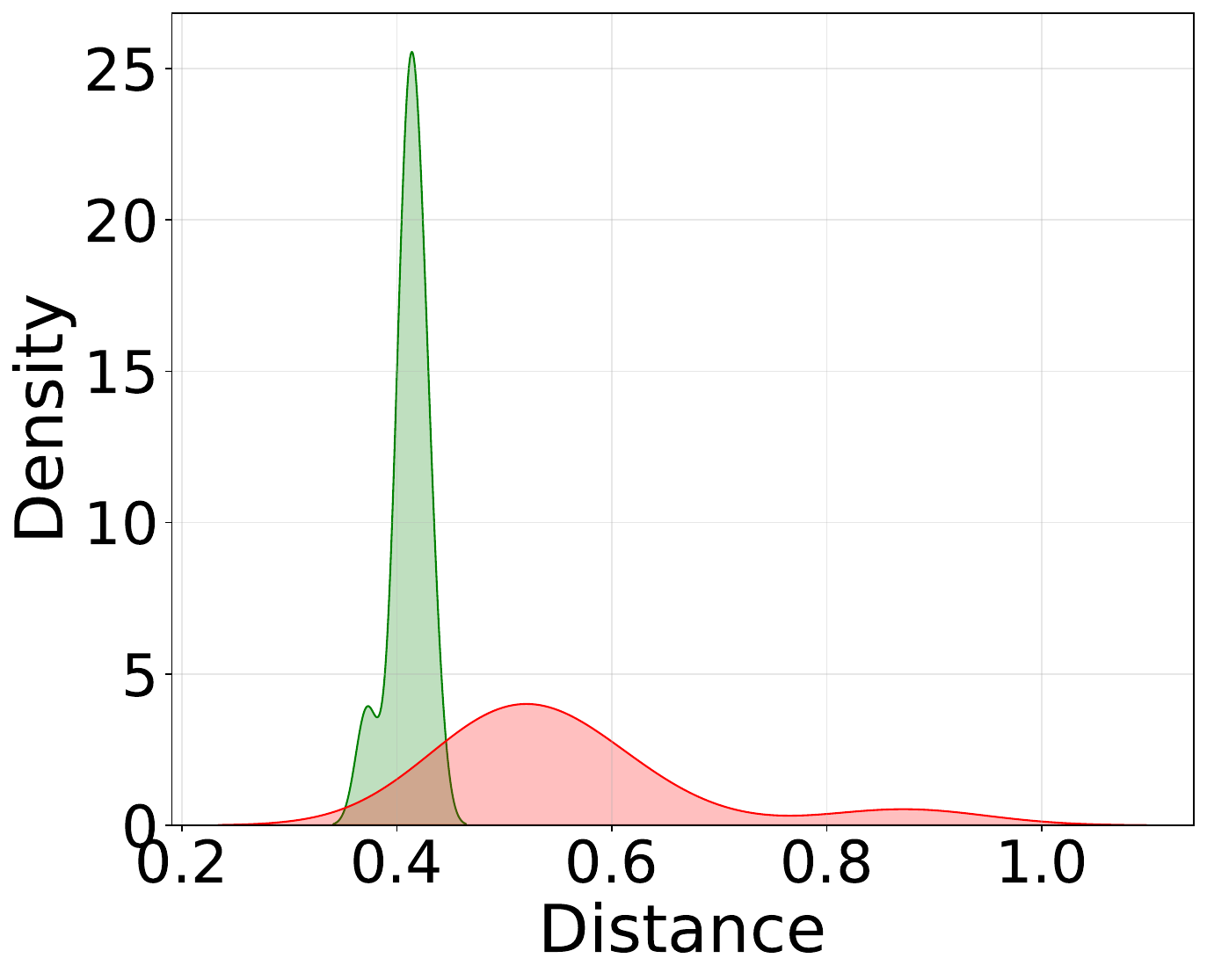}
        \caption{Refool}
    \end{subfigure}

    \medskip

    \begin{subfigure}{0.24\textwidth}
        \centering
        \includegraphics[width=\linewidth]{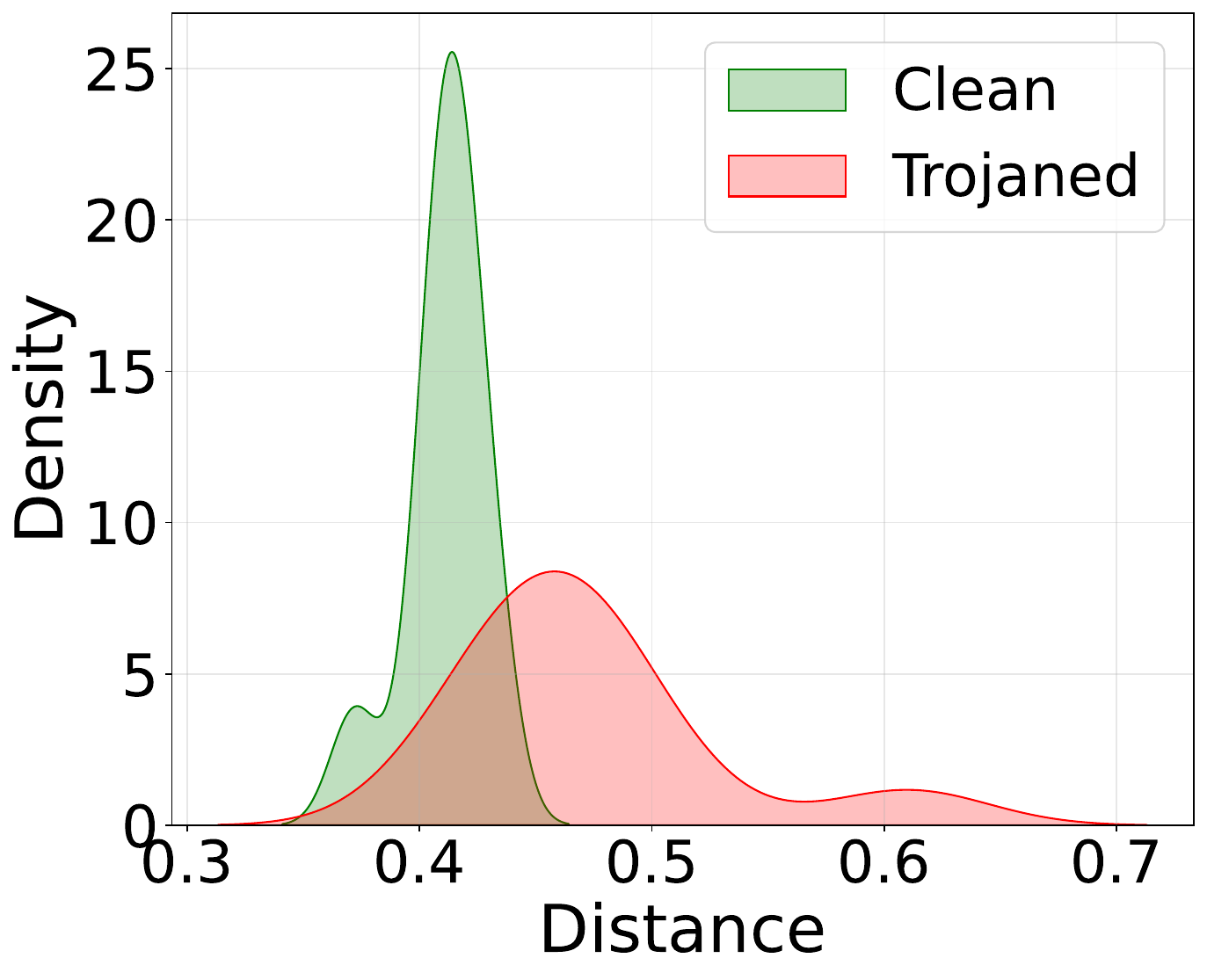}
        \caption{SIG}
    \end{subfigure}\hfill
    \begin{subfigure}{0.24\textwidth}
        \centering
        \includegraphics[width=\linewidth]{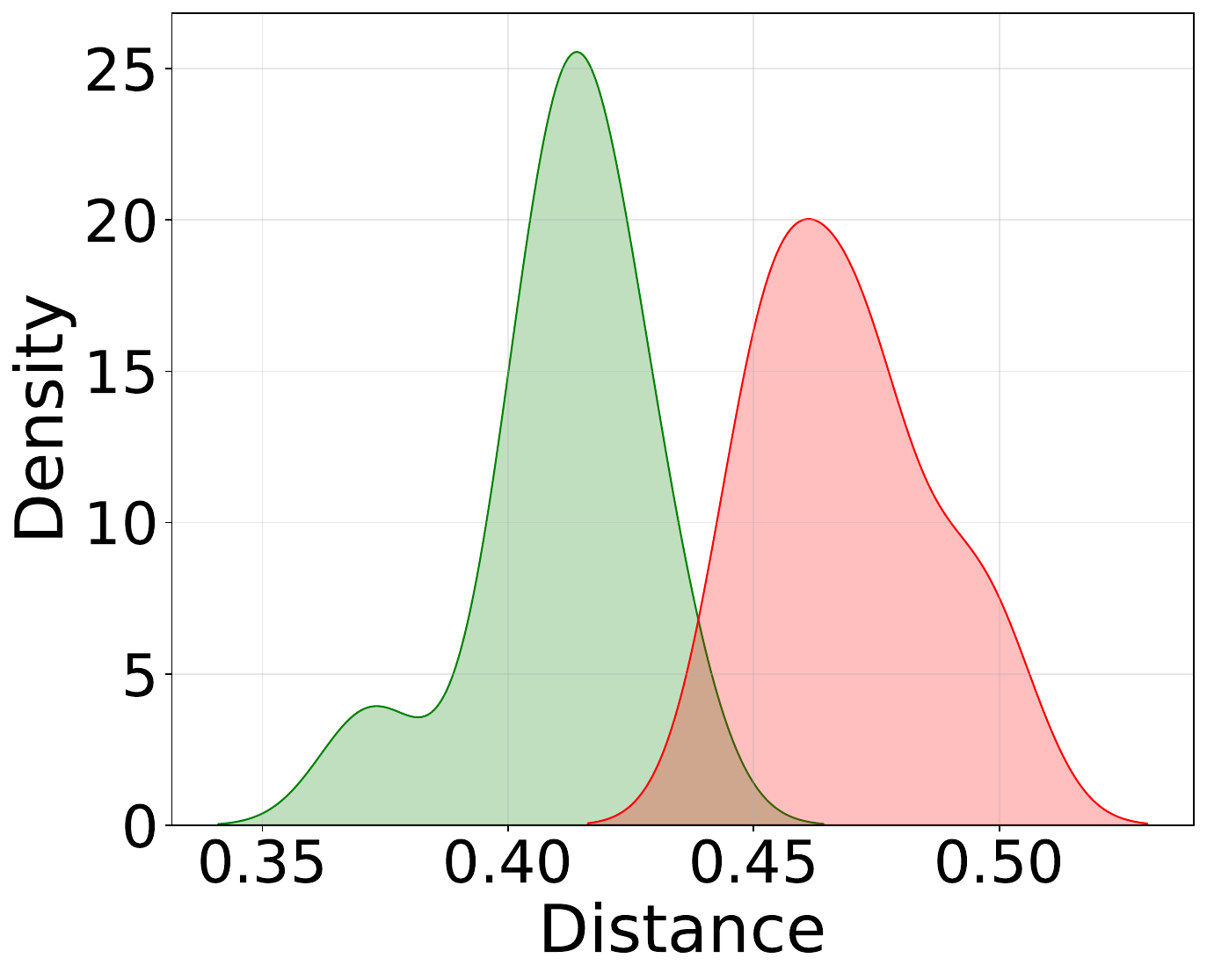}
        \caption{SSBA}
    \end{subfigure}\hfill
    \begin{subfigure}{0.24\textwidth}
        \centering
        \includegraphics[width=\linewidth]{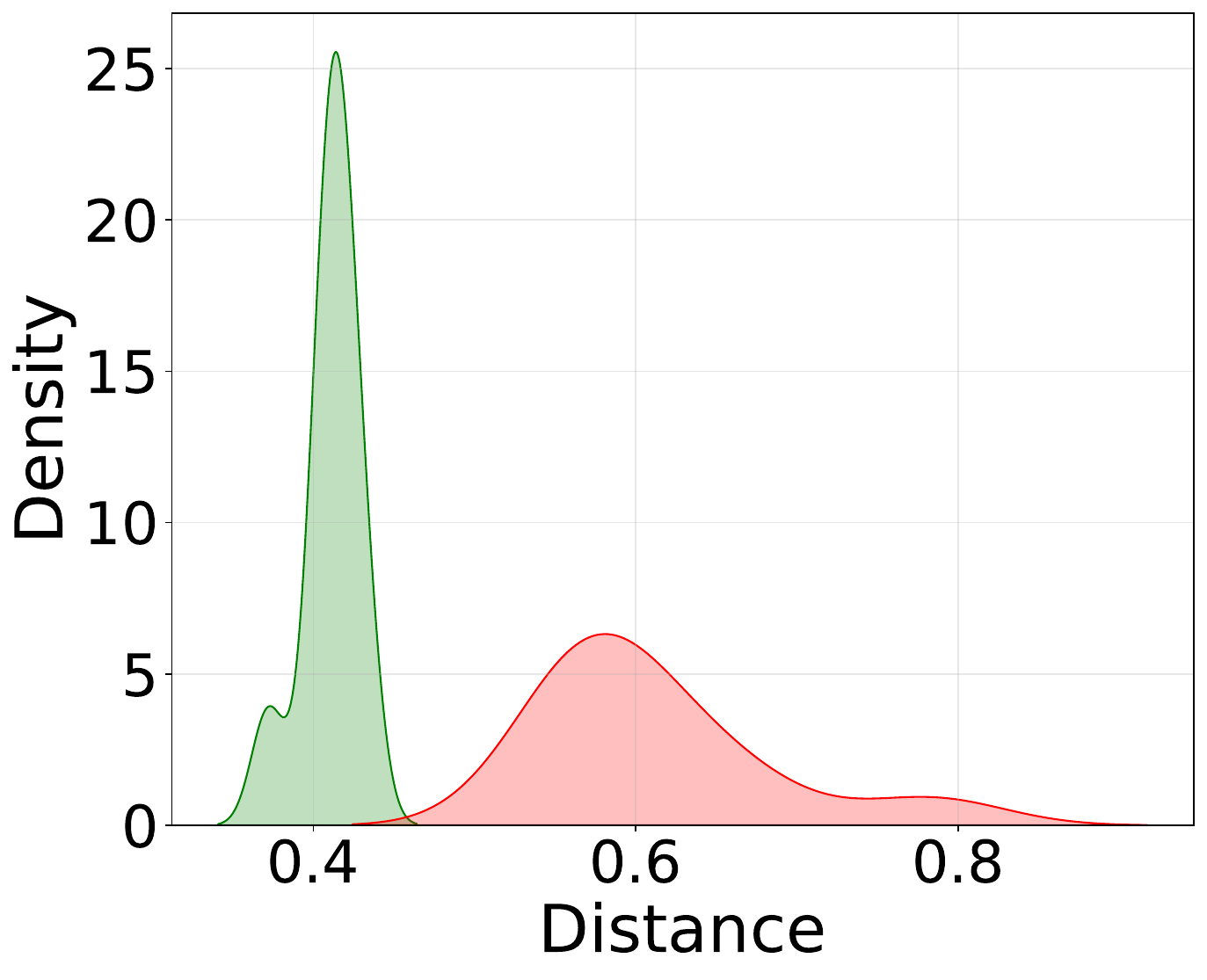}
        \caption{TrojanNN}
    \end{subfigure}\hfill
    \begin{subfigure}{0.24\textwidth}
        \centering
        \includegraphics[width=\linewidth]{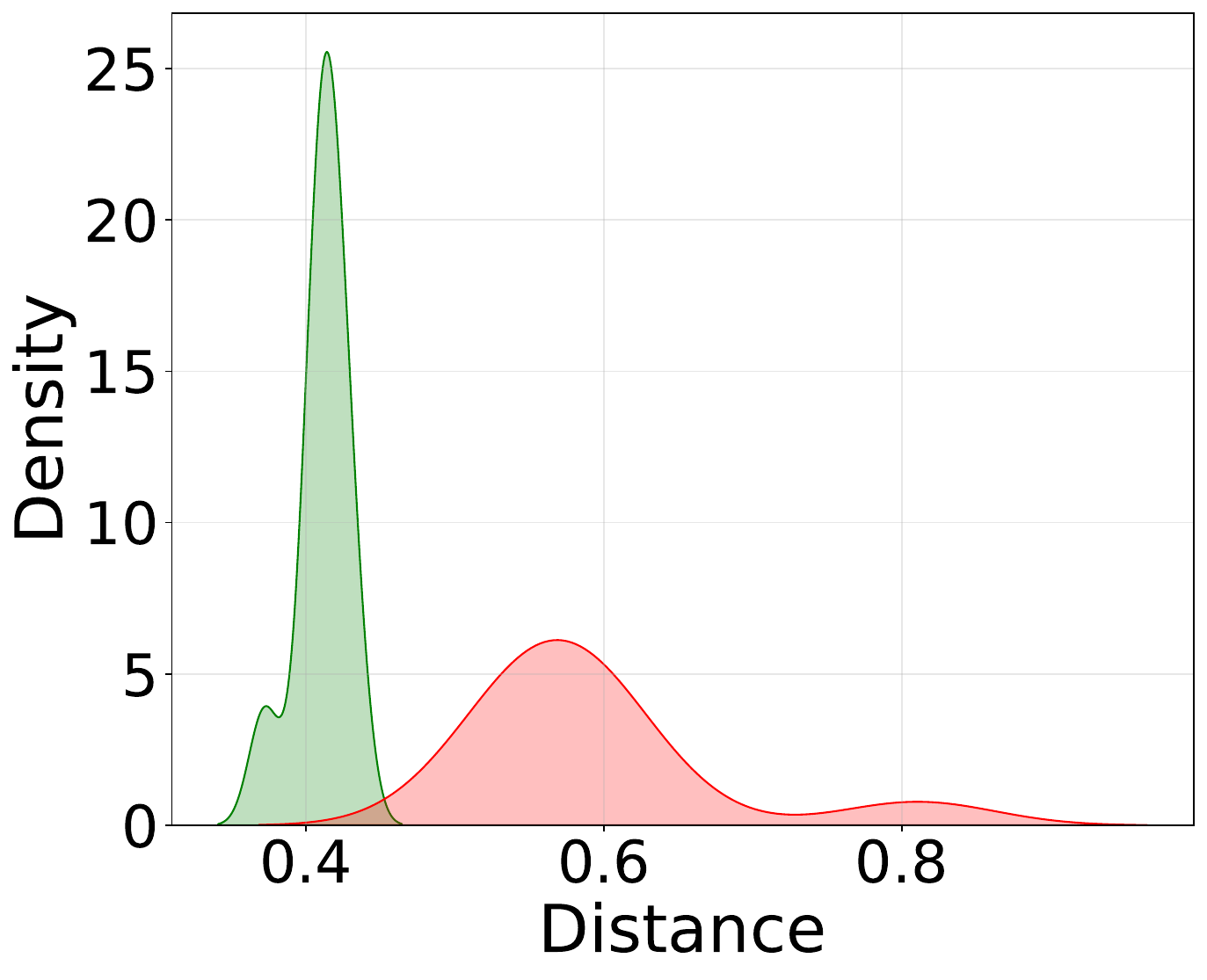}
        \caption{Wanet}
    \end{subfigure}

\caption{Distributions of spectral distances to the reference model $M^{0}$ for clean (green) and Trojaned (red) models on CIFAR-10. Each subfigure corresponds to a specific Trojan attack. Trojaned models consistently exhibit larger spectral distances, indicating separability from benign fine-tuned models.} %
    \label{fig:roc-aucs}
\end{figure}

Across attacks, the Blended triggers are the easiest to separate (average AUC 0.99), while SIG represents the most challenging case (average AUC 0.90). Importantly, no attack prevents separability across  datasets, suggesting that the observed effect is not tied to a specific trigger structure or attack strategy. Figure~\ref{fig:roc-aucs} provides a qualitative view of this phenomenon for CIFAR-10. For all attacks, the distributions of spectral distances for Trojaned models are consistently shifted toward larger values compared to clean fine-tuned models. While the degree of separation varies across attacks, the overall trend is stable: Trojaned updates induce spectral deviations that exceed those observed under benign evolution.

These results provide strong empirical support for the spectral separability assumption, showing that Trojan insertion leaves a measurable footprint in internal activation spectra, independent of the specific attack mechanism. This motivates the use of spectral distances as the basis for anomaly detection in the full \name pipeline.

\begin{tcolorbox}[boxrule=0pt,frame hidden,sharp corners,enhanced,borderline north={1pt}{0pt}{black},borderline south={1pt}{0pt}{black},boxsep=2pt,left=2pt,right=2pt,top=2.5pt,bottom=2pt]
\textbf{Answer to RQ1}:
Spectral distances reliably separate Trojaned models from clean fine-tuned models, achieving an average ROC--AUC of 0.95 across attacks and datasets. This confirms that Trojan insertion induces internal spectral changes that are distinct from those caused by benign model evolution.
\end{tcolorbox}

\subsection{Detection Effectiveness (RQ2)}

\begin{table*}[t]
\centering
\caption{Comparison of \name with state-of-the-art Trojan detectors. The highest accuracy for each dataset--attack pair is shown in \textbf{bold}. It is also \underline{underlined} when there is a statistically significant difference ($p < 0.05$) w.r.t. the second best approach on average (FeatuRE) according to the Fisher's exact test.}
\label{tab:rq2}
\resizebox{\textwidth}{!}{%
\begin{tabular}{ll rrrrr rrrrr rrrrr rrrrr}
\toprule
\multirow{2}{*}{\textbf{Dataset}} & \multirow{2}{*}{\textbf{Attack}}
& \multicolumn{5}{c}{\textbf{NC}}
& \multicolumn{5}{c}{\textbf{ABS}}
& \multicolumn{5}{c}{\textbf{FeatureRE}}
& \multicolumn{5}{c}{\textbf{\name}} \\
\cmidrule(lr){3-7}
\cmidrule(lr){8-12}
\cmidrule(lr){13-17}
\cmidrule(lr){18-22}
& 
& \textbf{TP} & \textbf{FP} & \textbf{FN} & \textbf{TN} & \textbf{Acc}
& \textbf{TP} & \textbf{FP} & \textbf{FN} & \textbf{TN} & \textbf{Acc}
& \textbf{TP} & \textbf{FP} & \textbf{FN} & \textbf{TN} & \textbf{Acc}
& \textbf{TP} & \textbf{FP} & \textbf{FN} & \textbf{TN} & \textbf{Acc} \\
\midrule
\multirow{8}{*}{CIFAR-10} & BadNet & 10 & 3 & 0 & 7 & 85\% & 7 & 2 & 3 & 8 & 75\% & 10 & 0 & 0 & 10 & \textbf{100\%} & 10 & 0 & 0 & 10 & \textbf{100\%} \\
 & Blended & 5 & 3 & 5 & 7 & 60\% & 5 & 2 & 5 & 8 & 65\% & 6 & 0 & 4 & 10 & 80\% & 10 & 0 & 0 & 10 & \textbf{100\%} \\
 & Input-Aware & 10 & 3 & 0 & 7 & 85\% & 6 & 2 & 4 & 8 & 70\% & 10 & 0 & 0 & 10 & \textbf{100\%} & 10 & 0 & 0 & 10 & \textbf{100\%} \\
 & Refool & 3 & 3 & 7 & 7 & 50\% & 7 & 2 & 3 & 8 & 75\% & 5 & 0 & 5 & 10 & 75\% & 10 & 0 & 0 & 10 &\underline{\textbf{ 100\%}} \\
 & SIG & 3 & 3 & 7 & 7 & 50\% & 5 & 2 & 5 & 8 & 65\% & 0 & 0 & 10 & 10 & 50\% & 7 & 0 & 3 & 10 & \underline{\textbf{85\%}} \\
 & SSBA & 9 & 3 & 1 & 7 & 80\% & 5 & 2 & 5 & 8 & 65\% & 1 & 0 & 9 & 10 & 55\% & 10 & 0 & 0 & 10 & \underline{\textbf{100\%}} \\
 & TrojanNN & 7 & 3 & 2 & 7 & 74\% & 6 & 2 & 4 & 8 & 70\% & 10 & 0 & 0 & 10 &\textbf{ 100\%} & 10 & 0 & 0 & 10 & \textbf{100\%} \\
 & Wanet & 4 & 3 & 6 & 7 & 55\% & 5 & 2 & 5 & 8 & 65\% & 10 & 0 & 0 & 10 & \textbf{100\%} & 10 & 0 & 0 & 10 & \textbf{100\%} \\
\midrule
\multirow{8}{*}{SVHN} & BadNet & 10 & 1 & 0 & 9 & 95\% & 3 & 0 & 7 & 10 & 65\% & 9 & 0 & 1 & 10 & 95\% & 10 & 0 & 0 & 10 & \textbf{100\%} \\
 & Blended & 6 & 1 & 4 & 9 & 75\% & 9 & 0 & 1 & 10 & \textbf{95\%} & 1 & 0 & 9 & 10 & 55\% & 9 & 0 & 1 & 10 & \underline{\textbf{95\%}} \\
 & Input-Aware & 7 & 1 & 3 & 9 & 80\% & 6 & 0 & 4 & 10 & 80\% & 10 & 0 & 0 & 10 & \textbf{100\%} & 10 & 0 & 0 & 10 & \textbf{100\%} \\
 & Refool & 2 & 1 & 8 & 9 & 55\% & 4 & 0 & 6 & 10 & 70\% & 6 & 0 & 3 & 10 & 84\% & 9 & 0 & 1 & 10 & \textbf{95\%} \\
 & SIG & 4 & 1 & 6 & 9 & 65\% & 6 & 0 & 4 & 10 & \textbf{80\%} & 0 & 0 & 9 & 10 & 53\% & 6 & 0 & 4 & 10 & \textbf{80\%} \\
 & SSBA & 9 & 1 & 1 & 9 & \textbf{90\%} & 6 & 0 & 4 & 10 & 80\% & 7 & 0 & 2 & 10 & 89\% & 8 & 0 & 2 & 10 & \textbf{90\%} \\
 & TrojanNN & 4 & 1 & 6 & 9 & 65\% & 7 & 0 & 3 & 10 & 85\% & 0 & 0 & 9 & 10 & 53\% & 10 & 0 & 0 & 10 & \underline{\textbf{100\%}} \\
 & Wanet & 8 & 1 & 2 & 9 & 85\% & 5 & 0 & 5 & 10 & 75\% & 9 & 0 & 0 & 10 & \textbf{100\%} & 10 & 0 & 0 & 10 & \textbf{100\%} \\
\midrule
\multirow{8}{*}{GTSRB} & BadNet & 10 & 7 & 0 & 3 & 65\% & 7 & 0 & 3 & 10 & 85\% & 5 & 2 & 5 & 8 & 65\% & 10 & 0 & 0 & 10 & \underline{\textbf{100\%}} \\
 & Blended & 10 & 7 & 0 & 3 & 65\% & 5 & 0 & 5 & 10 & 75\% & 4 & 2 & 6 & 8 & 60\% & 10 & 0 & 0 & 10 & \underline{\textbf{100\%}} \\
 & Input-Aware & 10 & 7 & 0 & 3 & 65\% & 5 & 0 & 5 & 10 & 75\% & 10 & 2 & 0 & 8 & 90\% & 10 & 0 & 0 & 10 & \textbf{100\%} \\
 & Refool & 5 & 7 & 5 & 3 & 40\% & 3 & 0 & 7 & 10 & 65\% & 0 & 2 & 10 & 8 & 40\% & 10 & 0 & 0 & 10 & \underline{\textbf{100\%}} \\
 & SIG & 6 & 7 & 4 & 3 & 45\% & 4 & 0 & 6 & 10 & 70\% & 10 & 2 & 0 & 8 & 90\% & 10 & 0 & 0 & 10 & \textbf{100\%} \\
 & SSBA & 8 & 7 & 2 & 3 & 55\% & 6 & 0 & 4 & 10 & 80\% & 0 & 2 & 10 & 8 & 40\% & 10 & 0 & 0 & 10 & \underline{\textbf{100\%}} \\
 & TrojanNN & 10 & 7 & 0 & 3 & 65\% & 4 & 0 & 6 & 10 & 70\% & 6 & 2 & 4 & 8 & 70\% & 10 & 0 & 0 & 10 & \underline{\textbf{100\%}} \\
 & Wanet & 4 & 7 & 6 & 3 & 35\% & 6 & 0 & 4 & 10 & 80\% & 10 & 2 & 0 & 8 & 90\% & 10 & 0 & 0 & 10 & \textbf{100\%} \\
\midrule
\multirow{8}{*}{CelebA} & BadNet & 4 & 1 & 6 & 9 & 65\% & 5 & 1 & 5 & 9 & 70\% & 10 & 4 & 0 & 6 & 80\% & 7 & 0 & 3 & 10 & \textbf{85\%} \\
 & Blended & 2 & 1 & 8 & 9 & 55\% & 6 & 1 & 4 & 9 & 75\% & 3 & 4 & 7 & 6 & 45\% & 8 & 0 & 2 & 10 & \underline{\textbf{90\%}} \\
 & Input-Aware & 4 & 1 & 6 & 9 & 65\% & 7 & 1 & 3 & 9 & 80\% & 10 & 4 & 0 & 6 & 80\% & 7 & 0 & 3 & 10 & \textbf{85\%} \\
 & Refool & 1 & 1 & 9 & 9 & 50\% & 4 & 1 & 6 & 9 & 65\% & 6 & 4 & 4 & 6 & 60\% & 7 & 0 & 3 & 10 & \textbf{85\%} \\
 & SIG & 0 & 1 & 10 & 9 & 45\% & 5 & 1 & 5 & 9 & 70\% & 6 & 4 & 4 & 6 & 60\% & 7 & 0 & 3 & 10 & \textbf{85\%} \\
 & SSBA & 1 & 1 & 9 & 9 & 50\% & 6 & 1 & 4 & 9 & 75\% & 9 & 4 & 1 & 6 & 75\% & 7 & 0 & 3 & 10 & \textbf{85\%} \\
 & TrojanNN & 7 & 1 & 3 & 9 & 80\% & 3 & 1 & 7 & 9 & 60\% & 9 & 4 & 1 & 6 & 75\% & 9 & 0 & 1 & 10 &\textbf{ 95\%} \\
 & Wanet & 2 & 1 & 8 & 9 & 55\% & 2 & 1 & 8 & 9 & 55\% & 10 & 4 & 0 & 6 & \textbf{80\%} & 5 & 0 & 5 & 10 & 75\% \\
\bottomrule
\end{tabular}
}
\end{table*}

RQ2 evaluates \name as a Trojan detector in its intended operational setting: detecting malicious updates after a single fine-tuning step. Table~\ref{tab:rq2} reports detection results across datasets and attacks, comparing \name against representative state-of-the-art Trojan detectors.

Overall, \name achieves an average detection accuracy of 0.95 across all attack--dataset combinations, substantially outperforming the baselines. Among the competing approaches, FeatuRE ranks second with an average accuracy of 0.75, followed by ABS (0.73) and Neural Cleanse (0.64). This performance gap indicates that the spectral signals identified in RQ1 translate into effective end-to-end detection capability. Across datasets, \name consistently achieves high accuracy: 0.98 on CIFAR-10, 0.95 on SVHN, 1.00 on GTSRB, and 0.86 on CelebA. These trends closely mirror the separability results observed in RQ1, reinforcing the connection between spectral separability and detection effectiveness.

Compared to FeatuRE, the strongest baseline, \name improves average detection accuracy by 20 percentage points\footnote{\textit{Percentage points} are the standard unit of measure for differences between percentages.}, with consistent gains of 16 points on CIFAR-10, 16 points on SVHN, 32 points on GTSRB, and 16 points on CelebA. This improvement is notable given that FeatuRE represents the most advanced form of trigger reverse engineering, operating in the feature space rather than the input space, and suggests that explicitly modeling benign spectral evolution provides a more robust detection signal than reconstructing Trojan-specific mechanisms.

Beyond overall accuracy, Table~\ref{tab:rq2} reveals systematic differences in error behavior. Baseline methods predominantly fail by missing Trojaned models (false negatives), particularly for attacks with distributed or input-dependent triggers such as SIG, Refool, and WaNet, and in addition incur non-negligible false-positive rates. In contrast, \name exhibits a more asymmetric error profile: false positives are rare, and detection failures, when they occur, are primarily false negatives. This low false-positive rates of \name indicate that it can be used as a validation step for model updates without frequently rejecting benign updates, a property that is not shared by the baseline methods.

Detection difficulty varies across attacks. TrojanNN is the easiest to detect, with an average accuracy of 0.99, whereas SIG poses the greatest challenge, with an average accuracy of 0.88. Despite this variability, \name consistently outperforms existing detectors. Across the 32 evaluated attack--dataset combinations, \name achieves strictly higher accuracy than all baselines in 24 cases, matches the best-performing baseline in 7 cases, and is outperformed only once, where FeatuRE exceeds \name by 5 percentage points.

We compared the best performing approaches (in bold in Table~\ref{tab:rq2}) with the second best performing approach on average (FeatuRE) using the Fisher's exact test, which compares the number of correct vs incorrect classifications made by the two approaches. When the $p$-value is lower than the standard 0.05 threshold, we underline the boldface value, to indicate statistical significance of the difference with the second best. 
We can notice that in 11 cases, not only \name achieves the best results, but it does so with a statistically significant margin over FeatuRE. The only case in which FeatuRE exceeds MIST is not statistically significant.
The same test was also executed to compare the aggregated results of MIST on all the Dataset-Attack pair against FeatuRE, and it resulted significantly better with a $p$-value of $9.35 \cdot 10^{-25}$.
These results should be interpreted in light of the assumptions underlying \name. Unlike baseline methods that aim to detect Trojans in general trained models, \name assumes access to a clean reference checkpoint and targets scenarios in which models are validated after fine-tuning against this reference. While this restricts applicability to a more specific deployment setting, it enables a focused analysis of model evolution. Within this setting, the results indicate that \name provides a consistently stronger and more stable detection signal than existing approaches.

\begin{tcolorbox}[boxrule=0pt,frame hidden,sharp corners,enhanced,borderline north={1pt}{0pt}{black},borderline south={1pt}{0pt}{black},boxsep=2pt,left=2pt,right=2pt,top=2.5pt,bottom=2pt]
\textbf{Answer to RQ2}: \name detects Trojaned model updates reliably after a single fine-tuning step, achieving an average detection accuracy of 0.95 across datasets and attacks, and consistently outperforming existing Trojan detectors. The results show that spectral deviations induced by malicious updates translate into a strong and stable detection signal in practice. Compared to trigger reconstruction-based approaches, \name exhibits fewer false positives and more predictable failure modes, making it suitable for validating model updates against a clean reference checkpoint.

\end{tcolorbox}

\subsection{Robustness under Multi-step Evolution (RQ3)}

\begin{table*}[t]
\centering
\caption{Multi-step performance of \name}
\label{tab:rq3}
\resizebox{\textwidth}{!}{%
\begin{tabular}{l rrrrr rrrrr rrrrr rrrrr}
\toprule
\multirow{2}{*}{\textbf{Attack}}
& \multicolumn{5}{c}{\textbf{CIFAR-10}}
& \multicolumn{5}{c}{\textbf{SVHN}}
& \multicolumn{5}{c}{\textbf{GTSRB}}
& \multicolumn{5}{c}{\textbf{CelebA}} \\
\cmidrule(lr){2-6}
\cmidrule(lr){7-11}
\cmidrule(lr){12-16}
\cmidrule(lr){17-21}
& \textbf{TP} & \textbf{FP} & \textbf{FN} & \textbf{TN} & \textbf{Acc}
& \textbf{TP} & \textbf{FP} & \textbf{FN} & \textbf{TN} & \textbf{Acc}
& \textbf{TP} & \textbf{FP} & \textbf{FN} & \textbf{TN} & \textbf{Acc}
& \textbf{TP} & \textbf{FP} & \textbf{FN} & \textbf{TN} & \textbf{Acc} \\
\midrule
BadNet
& 10 & 9 & 1 & 0 & 95\%
& 10 & 8 & 2 & 0 & 90\%
& 10 & 8 & 2 & 0 & 90\%
& 7 & 9 & 1 & 3 & 80\% \\
Blended
& 10 & 9 & 1 & 0 & 95\%
& 10 & 8 & 2 & 0 & 90\%
& 10 & 8 & 2 & 0 & 90\%
& 9 & 9 & 1 & 1 & 90\% \\
Input-Aware
& 10 & 9 & 1 & 0 & 95\%
& 10 & 8 & 2 & 0 & 90\%
& 10 & 8 & 2 & 0 & 90\%
& 6 & 9 & 1 & 4 & 75\% \\
Refool
& 10 & 9 & 1 & 0 & 95\%
& 10 & 8 & 2 & 0 & 90\%
& 10 & 8 & 2 & 0 & 90\%
& 8 & 9 & 1 & 2 & 85\% \\
SIG
& 7 & 9 & 1 & 3 & 80\%
& 7 & 8 & 2 & 3 & 75\%
& 10 & 8 & 2 & 0 & 90\%
& 7 & 9 & 1 & 3 & 90\% \\
SSBA
& 9 & 9 & 1 & 1 & 95\%
& 10 & 8 & 2 & 0 & 90\%
& 10 & 8 & 2 & 0 & 90\%
& 9 & 9 & 1 & 1 & 90\% \\
TrojanNN
& 10 & 9 & 1 & 0 & 95\%
& 10 & 8 & 2 & 0 & 90\%
& 10 & 8 & 2 & 0 & 90\%
& 10 & 9 & 1 & 0 & 95\% \\
Wanet
& 10 & 9 & 1 & 0 & 95\%
& 10 & 8 & 2 & 0 & 90\%
& 10 & 8 & 2 & 0 & 90\%
& 8 & 9 & 1 & 2 & 85\% \\
\bottomrule
\end{tabular}
}
\end{table*}

RQ3 examines whether \name remains effective as a model undergoes repeated benign evolution. In practice, deployed models are often fine-tuned multiple times, and deviations introduced by successive updates may accumulate. To evaluate robustness under such conditions, we reuse the CSDD computed for RQ2 and introduce an additional clean fine-tuning step before applying anomaly detection.

Table~\ref{tab:rq3} reports detection results across datasets and attacks. Overall, \name achieves an average accuracy of 0.89 in this multi-step setting. Performance remains high on CIFAR-10 (0.93), SVHN (0.88), and GTSRB (0.90), with a more noticeable reduction on CelebA (0.85), which represents the most challenging dataset in our evaluation. These results indicate that \name continues to provide meaningful detection signals even as the model evolves beyond a single update.

Compared to the single-step scenario in RQ2 (see Table~\ref{tab:rq2}), accuracy decreases by an average of 6 percentage points. This reduction is consistent across datasets and is not associated with specific attack types. Instead, inspection of the confusion matrices shows that the degradation is primarily due to an increase in \textit{false positives}, while true positive rates remain largely stable. In other words, repeated benign fine-tuning gradually changes the model relative to the original reference, leading to more false alarms, but Trojaned models continue to exhibit additional spectral deviations that remain detectable. This result suggests a predictable and bounded form of degradation. As the distance from the original reference checkpoint increases, \name becomes more conservative, occasionally flagging benign updates, but does not lose its ability to detect Trojaned models. In deployment settings where clean data becomes available after several updates, recomputing the CSDD can realign the reference distribution and mitigate this effect.

\begin{tcolorbox}[boxrule=0pt,frame hidden,sharp corners,enhanced,borderline north={1pt}{0pt}{black},borderline south={1pt}{0pt}{black},boxsep=2pt,left=2pt,right=2pt,top=2.5pt,bottom=2pt]
\textbf{Answer to RQ3}: \name remains effective under multi-step benign model evolution, achieving an average detection accuracy of 0.89. Performance degrades gracefully as models drift from the original reference, with degradation driven primarily by increased false positives rather than missed Trojans, indicating robustness to moderate levels of benign evolution.
\end{tcolorbox}

%% file: threats.tex
\section{Threats to Validity}
\label{sec:threats}

\paragraph{Conclusion validity}
Our experiments involve training and fine-tuning DNNs, which are inherently subject to non-determinism due to stochastic optimization. To mitigate this threat, we fixed random seeds and repeated the experiments multiple times. We checked the statistical significance of the differences using Fisher's exact test.

\paragraph{Internal validity}
Several baseline detectors required adapting publicly available implementations to support additional datasets and experimental settings. Although we followed the original descriptions and validated the adapted implementations through consistency checks, implementation-level differences may still affect the absolute performance.

\paragraph{External validity}
Our evaluation considers eight representative Trojan attacks, three state-of-the-art detectors, and four widely used datasets. While these choices cover a diverse range of attack mechanisms and data characteristics, they do not exhaust the space of possible Trojans, training procedures, or deployment scenarios. In particular, \name assumes access to a clean reference checkpoint and targets settings in which models are validated after fine-tuning. Detection performance may differ in scenarios that violate these assumptions. Moreover, our multi-step analysis evaluates robustness under a limited number of benign fine-tuning steps. As discussed in Section~\ref{sec:results}, repeated model evolution may require recomputation of the reference distribution when clean data becomes available, and behavior under long sequences of updates remains an open question.

\paragraph{Construct validity}
We evaluate detection performance using ROC--AUC and accuracy, which are standard metrics in the Trojan and anomaly detection literature. While these metrics capture overall separability and classification performance, they do not fully characterize operational trade-offs, such as the relative cost of false positives and false negatives in specific deployment settings. We therefore also report confusion matrix statistics to support a more detailed interpretation of detector behavior.

%% file: conclusion.tex
\section{Conclusion}
\label{sec:conclusion}

This paper presented \name, a Trojan detection approach that treats model updates as regression events and assesses whether fine-tuning introduces spectral changes consistent with benign evolution. Unlike prior work based on trigger reconstruction or input manipulation, \name leverages activation spectra as a compact and model-intrinsic representation of update behavior. We evaluated \name across four datasets and compared it against three state-of-the-art Trojan detectors. The results show that spectral distances reliably separate Trojaned updates from clean fine-tuning, achieving an average detection accuracy of 0.95 after a single update. Under multi-step benign evolution, \name remains effective, with average accuracy remaining at 0.89; although repeated fine-tuning introduces moderate benign drift, detection performance degrades gracefully and remains bounded.